\input{psfig}
\documentstyle[11pt]{uiucthesis}
\title{Cellular Games}
\author{Lenore E. Levine}
\degreeyear{1994}	
\phdthesis
\schools{B.S., State University of New York at Albany, 1988\\
M.S., University of Illinois, 1991}
\begin{document}
\copyrightpage
\maketitle

\begin{preliminary}
\begin{abstract}
A {\sl cellular game} is a dynamical system in which
cells, placed in some discrete structure, are regarded as playing
a game with their immediate neighbors. Individual strategies may
be either deterministic or stochastic. Strategy success is
measured according to some universal and unchanging criterion.
Successful strategies persist and spread; unsuccessful ones
disappear.

In this thesis, two cellular game models are formally defined, and
are compared to cellular automata. Computer simulations of these
models are presented.

Conditions providing maximal average cell success, on one and
two-dimensional lattices, are examined. It is shown that these
conditions are not necessarily stable; and an example of such
instability is analyzed. It is also shown that Nash equilibrium
strategies are not necessarily stable.

Finally, a particular kind of zero-depth, two-strategy cellular
game is discussed; such a game is called a {\sl simple cellular
game}. It is shown that if a simple cellular game is left/right
symmetric, and if there are initially only finitely
many cells using one strategy, the zone in which this strategy
occurs has probability $0$ of expanding arbitrarily far
in one direction only. With probability $1$, it will either
expand in both directions or disappear.

Computer simulations of such games are presented. These
experiments suggest the existence of two different kinds of
asymptotic behavior.
\vfill
\end{abstract}

\begin{dedication}
To My Mother, Dinah Green Levine
\end{dedication}

\begin{acknowledgements}
I would like to thank my advisor, Julian Palmore, for his
guidance and support. I would also like to thank Norman Packard for
introducing me to this new and challenging area, and
Larry Dornhoff for help with the computers.

In addition, I would like to thank the faculty of the
UIUC Department of Mathematics -- particularly Felix Albrecht,
Stephanie Alexander, Robert Muncaster,
Jerry Uhl and Wilson Zaring -- for their encouragement with my studies.
I would like to thank my roommates for their patience. And finally,
I would like to thank Roberta Hatch and ``A.T.'' for providing their
own form of inspiration.
\end{acknowledgements}


\tableofcontents
\listoffigures

\end{preliminary}

\begin{thesis}

\chapter{Introduction}

A {\sl cellular game} is a dynamical system; that is, the
variables it is composed of are regarded as changing
over time. These variables or cells,
arranged in a discrete structure such as a ring, are thought of as
repeatedly playing a game with their neighbors. Most of this
paper is concerned with one-dimensional cellular games, defined
more formally as follows:

\newtheorem {defn} {Definition} [chapter]
\begin{defn}
\label{cgame}
A {\bf one-dimensional cellular game} consists of:
\begin{enumerate}
\item
A one-dimensional discrete structure, uniform from the viewpoint of
each site; that is, a ring or doubly infinite path.
\item
A variable, or {\bf cell}, at each site. The components of
this variable may change at each discrete unit of time, or
{\bf round}. They consist of, at least:
\begin{enumerate}
\item A {\bf move} component, which can take on a finite number $k$
of values.
\item A {\bf strategy} component, which determines what
move a cell makes in a given round. The strategy of a cell
is based on past moves of it and its $r$ nearest neighbors
on each side. The number of past rounds considered is called the
{\bf depth} $d$
of the strategy. This $r$, as used above, is the {\bf radius}
of the game.
\end{enumerate}
\item
A fitness criterion, which does not change and is the same
for each cell. This fitness criterion is usually local; that
is, the fitness of a cell in each round is based on its move,
and those of nearest neighbors within the radius of the game.
\item
A mechanism for strategy selection, under which more fit
strategies survive and spread. Strategy selection is
usually nonlocal; that is, a more fit strategy may
spread arbitrarily far in a fixed number of time units.
An interval between strategy changes, which may be one or more rounds,
is called a {\bf generation}.
\end{enumerate}
\end{defn}

Thus, a cellular game can be considered a process in which cells
make moves each round, based on their strategies, and strategies
are updated in each generation, based on their fitness in
preceding rounds.

Note that cellular game strategies and fitness criteria
are usually stored in the form of a table. Also note
that $n$-dimensional cellular automata, with one cell for
each $n$-tuple of integers or integers mod $k$, can be
similarly defined.

One-dimensional cellular games are studied in
\cite {rogers}, \cite{cowan}, \cite{cowan2}, and \cite{miller}.
Similar systems are discussed in \cite{mat1}, \cite{mat2} and
\cite{mat3}; and games on a two-dimensional lattice in
\cite{nowak}.

Cellular games satisfy a criterion for ``artificial
life'' as discussed by Christopher Langton \cite{langton}. That
is, ``There are {\sl no} rules in the system that dictate global
behavior. Any behavior at levels higher than the (individual
cells) is, therefore, emergent.''

Cellular games are a generalization and extension of another,
more well-known, discrete dynamical system; that is, of {\sl
cellular automata}. They were created largely because of
questions arising from the observation of cellular automata.
One-dimensional cellular automata are defined as follows:

\label{CA}
\begin{defn}
A {\bf one-dimensional cellular automaton} con\-sists of:
\begin{itemize}
\item
A one-dimensional discrete structure, uniform from the viewpoint of
each site; that is, a ring or doubly infinite path.
\item
A variable, or {\bf cell}, at each site, that can
take on finitely many values or {\bf states}. The initial
states of a cell may be specified as desired.
\item A function which decides how each cell changes state
from one {\bf generation}, or discrete unit of time, to the next.
This function, or {\bf cellular
automaton rule}, is always the same for each cell, and
depends entirely on the state of a cell and that of its $r$ neighbors
on each side in the past $m$ generations. This $r$ is referred to
as the {\bf radius} of the cellular automaton, and $m$ as its
{\bf order}. Cellular automaton rules are usually stored
and described in the form of a table.
\end{itemize}
\end{defn}

It can be shown that a $m$th-order cellular automaton
is equivalent to a first-order cellular automaton with more
states. This proof \cite{waterman}, however, is dependent on
the locality of cellular automata -- that is, on the
fact that cells are directly affected only by their
neighbors. For similar mathematical objects, such as cellular
games, that are {\it not} local, this proof cannot be used.

Thus, if a cellular automaton, of radius $r$, operates on cells
that can take $k$ possible states, there are $k^{2r+1}$ possible
circumstances that need to be considered. The rule table,
therefore, has $k^{2r+1}$ entries; and there are $k^{k^{2r+1}}$
possible $r$-radius, $k$-state cellular automaton rules.
An example of a cellular automaton rule is the two-state, radius
one rule whose evolution is illustrated below. In this rule, a
cell can be in either state $0$ or state $1$. Any cell that, in
generation $g$ is in state $1$, and has both of its neighbors in
state $1$, stays in state $1$ in generation $g+1$. Otherwise, a
cell is in state $0$ in generation $g+1$. This rule is Rule 128
according to Wolfram's \cite {wolfram} classification system of
the $256$ $2$-state, radius one rules.

\begin{tabbing}
Generation 1:xx\=
1 xx\=0 xx\= 1 xx\= 1 xx\= 1 xx\= 1 xx\= 0 xx\= 1 xx\= 0xx\= 1 xx\kill
Generation 1:\> 1\> 0\> 1\> 1\> 1\> 1\> 0\> 1\> 0\> 1\\
Generation 2:\> 0\> 0\> 0\> 1\> 1\> 0\> 0\> 0\> 0\> 0\\
Generation 3:\> 0\> 0\> 0\> 0\> 0\> 0\> 0\> 0\> 0\> 0\\
\end{tabbing}

\newtheorem{tabl} [defn] {Table}
\begin{tabl}
The action of rule 128 on a circular ring of ten cells, for
three generations.
\end{tabl}

\begin{defn}
A {\bf stochastic cellular automaton} is as above, except that
neighborhood states do not determine the move made in
the next generation, but the probability that a particular
move will be made.
\end{defn}

Computer experiments on one-dimensional cellular automata are
usually conducted with cells arranged in a ring. Cell states are
indicated by colors; thus, $k$-state cellular automaton rules are
often referred to as $k$-color rules. Initial conditions are
displayed in a line on top of the screen, with each generation
being displayed below the previous generation. In such
experiments, initial conditions, and rule table entries, are
often chosen with the aid of a pseudorandom number generator.

As a matter of fact, descriptions of computer experiments with
cellular automata and other discrete dynamical systems often make
reference, informally, to ``random'' initial conditions. This
concept actually applies to mathematical models containing
infinitely many variables, such as a one-dimensional cellular
automaton with one cell for each integer. In such a case,
``random,'' ``almost all,'' or ``normal'' initial conditions
refer to conditions such that all $k^n$ of the $n$-tuples of $k$ cell
states are equally likely, for all $n$. Or, in other words, if
the states of the cells are construed as decimal places of two
real numbers, both numbers are normal to base $k$.

Such conditions cannot be exactly duplicated in the finite
case, no matter how large the number of cells. However,
conditions can be created which appear disordered and satisfy
certain statistical tests of disorder. This is done with the aid
of a pseudorandom number generator. Such initial conditions are often
loosely referred to as ``random.'' Computer simulations of discrete
dynamical systems often use such initial conditions as the most
feasible indicator of likely behavior.

In such experiments, there are, roughly, three types of
asymptotic behavior. First of all, all cells
may become and remain one color, or change color periodically,
with a small and easily observable period. Second, cells may
display ``chaotic'' behavior; that is, cell color choice may
appear to be disordered, or to result from some other simple
stochastic algorithm. Third, cell color
choice may be neither periodic nor chaotic, but appear to display
organized complexity.
That is, the cell evolution diagrams may look like biological
structures, such as plants, or social structures, such as city
maps. As a matter of fact, such diagrams are often quite
esthetically pleasing. These rule types are discussed
in \cite {wolfram2}; for more on the concept of ``complexity,''
as it applies to cellular automaton rules, see \cite {waldrop}.

On a finite ring of cells, of course, all such evolution is
eventually periodic. But, if cells can be in $2$ states, and
there are $640$ cells, there are $2^{640}$ possible ring states.
Therefore the period of ring states could, conceivably, be quite
high; and ``chaotic'' or ``complex'' rules do indeed seem to have
very high periods.

Visual representations of cellular automata can exhibit a
sophistication reminiscent of living structures. However, the
number of $k$-state, $r$-radius cellular automaton rules is very
large ($k^{k^{2r+1}}$) for all but the smallest $k$ and $r$; and
``interesting'' rules are not common and difficult to find. This
leads to the question, therefore, of whether there is some way of
``evolving'' cellular automaton rules in a desired direction.

There are two possible avenues of approach to this question. One
is to select rules based on their global properties. That is,
some computable measure of the desired characteristics is
devised, and rules are chosen by
their ability to meet this measure. Such selections are discussed
in \cite{packard2} and \cite{mitchell}.

The other way is to select rules based on their local properties.
That is, each cell uses a different rule; and there is some
universal and unchanging criterion for rule success.
This approach is more like the way living systems evolve, for
the evolution of a planetary ecology is not due to
constraints placed directly on the ecology. It is an emergent
property of constraints placed on the individual organisms. For
this reason, such models may
potentially reveal not only the nature of ``complex'' rules,
but also how their global properties emerge from local
interactions.

An evolutionary model of this sort is equivalent to a cellular
game; the only difference is the terminology. That is,
the strategy of a cell can be regarded as the individual
rule used by each cell; the depth of the strategy
as the order of the rule; cell moves as states; and instead
of referring to the smallest unit of time as a round,
and a possibly larger unit as a generation, the smallest
unit can, as with cellular automata, be referred to
as a generation. The fitness criteria and evolutionary
process stay the same.

A cellular game differs from a cellular automaton not only in the
precise definition used, but also in the philosophy under which
this definition was constructed. That is:

\begin{itemize}
\item
Cellular automata are often regarded as a physical
models; for example, each cell may be seen as
an individual atom. Thus, the rules by which each
cell operates are the same. Cellular games, on the
other hand, are seen as an evolutionary models.
Each cell uses an individual rule, or strategy, which can be
thought of as the ``genetic code'' of the cell.
\item
Cellular automata are usually thought of as deterministic,
beyond the initial generation, though stochastic cellular automata
have also been studied. Cellular games operate stochastically;
that is, the evolutionary process under which strategies
are modified is stochastic, and, often, the strategies themselves
are sto\-chas\-tic.
\item
Cellular automata are local; that is, the state of a cell is
affected only by the states of its $r$ nearest neighbors
on each side in the previous generation. In other words,
cell information cannot travel more than $r$ units per
generation. This speed is often called ``the speed of light.''
Cellular games, on the other hand, typically use nonlocal
strategy selection criteria. That is, a more fit strategy may
propagate arbitrarily far in one generation. (There is
more discussion of the nonlocality of cellular games in
Section \ref{ZDM}.)
\item
In \cite{waterman}, it is shown that $m$th-order cellular automata
are behaviorally equivalent to first-order cellular automata
with larger radius and more states. However, this
proof does not work for cellular games with nonlocal
selection criteria. Moreover, cellular games are often
constructed with strategies looking more than one generation back.
\end{itemize}

Now, it can be shown that if a cellular game has a local
fitness criterion and local rule selection process,
it is actually equivalent to a cellular
automaton with a large number of states. This automaton, of
course, will be stochastic if the game is stochastic.

\newtheorem{thm} [defn] {Theorem}
\begin{thm}
Let $G$ be a cellular game with a local fitness criterion
and local rule selection process, which operates every $R$
rounds. Let all fitness measurements start over
again after this process. Then $G$ is equivalent to a cellular
automaton $G'$ with a much larger number of states.
\end{thm}

{\sl Proof.} Let $G'$ be constructed as follows:
let the state of a cell $c$ in $G'$ be a vector with the
following components:

\begin{enumerate}
\item
The state of $c$ in $G$.
\item
The individual rule used by $c$, in $G$.
\item
A $R$-valued counting variable, which starts out as $1$
in the first generation, and thereafter corresponds to
the current generation mod $R$.
\item
A fitness variable, which corresponds to the accumulated
fitness of a cell over $R$ rounds.
\end{enumerate}

Since these components enable $G'$ to simulate the action
of $G$, it suffices to show that $G'$ is a cellular automaton.
That is, each component must have only finitely many possible
values, and be locally determined. This is shown to be
true for each component, as follows:

\begin{enumerate}
\item
By definition of $G$, the first component has only finitely
many values. It is determined by the rule of a cell, and the
states of it and its neighbors in preceding rounds.
\item
By definition \ref{cgame}, even if stochastic rules are
used only finitely many are considered.
Whether or not a cell keeps its rule, after $R$ rounds,
is based on its own fitness, and the process of selecting
new rules is assumed to be local.
\item
The counting component can be in any one of $R$ different
states. The rule for its change is simple: If it is
in state $s$ in round $d$, it is in state $s+1$ $mod$ $g$
in round $d+1$. Note that to run $G'$ as a simulation of $G$,
this counting component must be initially set at the same
value for all cells.
\item
The fitness component is set to zero after every $R$
rounds; and can be incremented or decremented
in only finitely many different ways.
How it changes in each generation, for a given cell $c$, depends on the
first components of cells $c-r$ through $c+r$.
\end{enumerate}
\rule{2mm}{3mm}

Given this equivalence, why, then, is a cellular game so
different from a cellular automaton? For one thing, cellular games
often do use a nonlocal strategy selection process; it may
be considered an approximation to a selection process that
can operate over very large distances.
For another, cellular automaton rule spaces, especially those with high
radius, typically contain very large numbers of rules.
Therefore, even if only systems with a local selection process are
considered, the evolutionary paradigm of cellular games may still
be valuable. It may be a practical method of selecting members of
these spaces with interesting properties.

In this paper, two different models of cellular games are
defined. The original Arthur-Packard-Rogers model is discussed
first in Section \ref{APR}. This model is quite extensive and uses
many different parameters. The second, simplified, model is more
amenable to mathematical analysis. This model is discussed in
Section \ref{ZDM}.

Computer simulations of both models are presented. These
simulations are similar to those of cellular automata, both in
the way they are conducted and in the way they are displayed.
That is, cell moves are indicated by colors. Strategies are
usually not pictured, due to the large size of strategy spaces.
Thus, the move of a cell may also be referred to as its color. Initial
moves of a finite ring of cells are displayed in a line on top of
the screen, and each generation is displayed below the previous
generation. Initial moves and strategies, as well as other
stochastic choices during the course of the game, are implemented
with the aid of a pseudorandom number generator.

Computer simulations of the first model display sophisticated
behavior reminiscent of living systems, or ``complicated''
cellular automata. These behaviors, which include such phenomena
as zone growth and ``punctuated equilibria,'' are discussed and
extensively illustrated in Section \ref{comp1}.

The second model admits only deterministic strategies of depth
zero; that is, strategies of the form, ``Do move $m$, without
regard to previous rounds.'' Thus, in this model, moves and
strategies can be considered equivalent. Though this model is
simpler, there are still counterintuitive results associated with
it. Even if only two strategies are allowed under this model, it
is extremely difficult to predict which, if either, will be
stable under invasion by the other. There are no simple algorithms for
determining this.

For example, consider ring viability, discussed in
Section \ref{rvsec}. For finite rings this
concept, Definition \ref{ringviability}, refers
to the average success of all cells in the ring. In this chapter,
it is shown that under any local fitness criterion $G$,
rings in which the cells have made periodic move sequences have
the highest possible viability. It is also shown that a similar result
is false in the two-dimensional case.

Now, if cellular games did indeed always evolve towards highest
ring viability, this would make their course relatively easy to
predict. However, in Section \ref{stabsec}, a two-strategy
cellular game is presented, in
which the best strategy for the ring as a whole -- that is, the strategy
that, if every cell follows it, maximizes ring viability -- is not
stable under invasion. This instability is illustrated by
computer simulations, and is also proved. This is done by
showing that if a small number of cells using the invading
strategy are surrounded by large numbers that are not, the
invading strategy tends to spread in the next generation. The
reason for this is that the first strategy, though it does well
against itself, does poorly against the second one.

On the other hand, a winning strategy may not necessarily be
stable either. That is, strategy A may defeat strategy B, but
still be unable to resist invasion by it. The reason, in this
case, is that strategy B does so much better against itself. This
result can also be demonstrated by computer simulations and
proved, using the same method. These results are also in
Section \ref{stabsec}.

Finally, consider a situation in which, if its neighbors use strategy
A, a cell has greatest success if it uses strategy A too.
It seems logical that, in this case, strategy A would
indeed be stable. As a matter of fact, such a situation
is called, in game theory, a {\sl symmetric Nash equilibrium}.

However, it can be demonstrated by computer
simulations, and also proved, that some symmetric Nash equilibrium
strategies are {\sl not} stable under invasion. The reason, in
such cases, is that strategy B has somewhat less probability of
surviving in a strategy A environment, but is very good at
causing strategy A not to survive. Therefore strategy B is
somewhat less likely to persist, but is a lot more likely to spread.
This result is also considered in Section \ref{stabsec}.

Thus, the three theorems in Section \ref{stabsec} show how
difficult it is to predict the course of cellular games,
even under a very simple model. The counterintuitive
nature of the results obtained suggests the potential
mathematical interest of this paradigm.

The second part of this thesis presents results applicable
to particular examples of the zero-depth model, called
{\sl simple cellular games}. These games have two distinguishing
characteristics:
\begin{itemize}
\item{There are only two possible strategies; these two
strategies are referred to as white, and black.}
\item{Each cell has, at all times, positive probability of
either living or not living.}
\end{itemize}

The theorems discussed in the second part apply to simple
cellular games which are left/right symmetric. The Double Glider
Theorem, \ref {main}, applies to the evolution of such games
under initial conditions under which there are only
finitely many black cells. The {\sl zone of uncertainty}
is defined as the zone between the leftmost and rightmost
black cell. It is shown that the probability this zone
will expand arbitrarily far in one direction only is $0$.
That is, with probability $1$, it will either expand in
both directions or disappear.

Section \ref{stanrec}, which follows, discusses simple
game evolution in a slightly different context; that is, under conditions
such that there is a leftmost white cell and a rightmost black
cell, or {\sl standard restricted initial conditions}.
Simple cellular games with both left/right
and black/white symmetry are classified according to their
asymptotic behavior under these circumstances.
That is, they are divided into {\sl mixing processes}
and {\sl clumping processes}. The behavior of clumping processes
is further explored, and a conjecture is made that applies to
both kinds of processes.

In Section \ref{examples}, the last chapter, specific examples of simple
cellular games are presented. Computer simulations suggest
that one of these examples, the Join or Die Process, is a
clumping process; and the other, the Mixing Process, is,
as named, a mixing process.

\chapter{Cellular Game Models}
\label{CGM}
\section{Game Theory and Cellular Games}

Success criteria in tabular form, or score tables, are
extensively used in game theory. They describe the course of any
game which can be exactly modelled, for which strategy success
can be numerically described, and in which all strategies are
based on finite, exact information. For example, consider the
game of Scissors, Paper, Stone; that is, Scissors beats Paper,
Paper beats Stone, and Stone beats Scissors. Suppose this game is
played for one round, and the only possible strategies are
deterministic. Then the table for this game is (if a win scores
1, tie at .5 and loss at 0):

\begin{tabbing}
Opponent xxxx\= Sxxcissors\= xxPaper\= xxStone\kill
Opponent \>    Scissors \>   Paper \>   Stone \\
\\
Player \\
Scissors \>       .5  \>        1 \>      0 \\
Paper   \>        0   \>        .5 \>     1 \\
Stone   \>        1   \>        0  \>     .5 \\
\end{tabbing}

The following definition is used in game theory:

\begin{defn}
A {\bf mixed strategy} is a stochastic strategy; that is, one under
which, in some specified circumstances, more than one move has
positive probability.
\end{defn}

A table can also be devised for mixed strategies, and for games
of more than one round. For mixed strategies the table entry
describes the expected success.

For example, suppose the game of
Scissors, Paper Stone is played for two rounds, and there are
three possible strategies. Strategy A is to choose each move with
probability ${1 \over 3}$, Strategy B is to choose Stone for the
first move, and the move chosen by the other player for the
second, and Strategy C is always to choose Paper. Then the table
for this game is:

\begin{tabbing}
Strategy AAAC\= Strategy CA\= Strategy CA\= Strategy AC\kill
Opponent \>  Strategy A \>  Strategy B \>  Strategy C \\
\\
Player \\
Strategy A \>       1\>           1\>           1\\
Strategy B\>        1\>           1\>           .5\\
Strategy C\>        1\>          1.5\>          1\\
\end{tabbing}

\begin{defn}
A table depicting strategy success as described above is
called the {\bf normal form} of a game.
\end{defn}

Normal
form can be used, at least theoretically, to describe extremely
sophisticated games. For example, if only a fixed finite number
of moves are allowed, and strategies consider only the history of
the current game, then there are only finitely many deterministic
strategies for the game of chess. Hence normal form could,
at least theoretically, be used to describe this game.
Of course, there are so many possible chess strategies that this
form cannot be used for practical purposes. For more on
normal form, see \cite{luce}.

Note that this form is ambiguous if mixed strategies are allowed.
For example, consider the above table. Does it indicate the
actual success levels of deterministic strategies, or the
expected success levels of stochastic ones? It is not possible to
tell without further information.

Such a normal form can also be used to describe three-player
games. For example, this table describes a
game in which there are two moves, you score .85 if you make the
same move as both other players and .15 otherwise. This game
is called the Join or Die game.

\begin{tabbing}
Your Move: xx\=  B xx \= xx W exextra\= Your Move: xx\=   xx B\=   xx W\kill
Your Move:\> B\> \>                      Your Move:\> W\\
\\
Player 1:\>    B\>    W\>               Player 1:\>    B\>    W\\
\\
Player 2:\> \> \>                       Player 2:\\
   B \>        .85\>    .15\>                  B\>         .15\>    .15\\
   W \>        .15\>    .15\>                  W\>         .15\>    .85\\
\end{tabbing}

Now, consider cellular games. If the success criterion, or score, is
local; that is, if it
is based entirely on the state of a cell and those of its neighbors, it
can also be encoded as a table. As a matter of fact, any game
table for $2r + 1$ players can be used as the score table for a
cellular game of radius $r$. For example, the Join or Die process
is a cellular game of radius $1$, in which each cell plays the
Join or Die game with its two nearest neighbors. The following
table is used for this process:

\begin{tabbing}
Right Neighbor: xx\= B xx \= xx W exextra\=  Right Neighbor: xx\= xx B\= xx
W\kill
Cell's Move:\> B \> \>                   Cell's Move:\> W \\
\\
Right Neighbor: \>   B \>   W \>        Right Neighbor:\>    B\>    W\\
\\
Left Neighbor: \> \> \>                 Left Neighbor: \\
      B  \>          .85 \>   .15  \>             B \>           .15 \>   .15\\
      W  \>          .15 \>   .15  \>             W \>           .15 \>   .85\\
\end{tabbing}

However, cellular games differ from the situations most analyzed
by game theorists, or the vernacular notion of a game, in the
following ways:

\begin{itemize}
\item Each cell interacts with different neighbors, as determined
by the discrete structure on which the cellular game is run. That is,
the score of cell $0$ is based on its move, and those of cells $1$ and $-1$.
The score of cell $1$ is based on the moves of cells $0$ and $2$, not
cells $0$ and $-1$.

\item The ``game'' is considered to be played repeatedly, for
many rounds. Thus, the main focus is on optimal
move behavior in the long run, not for one round only.

\item There is an explicit mechanism for determining how successful
strategies thrive and spread. The
cellular game is not completely described without this mechanism;
no assumptions about asymptotic behavior can be made just on the
basis of the score table.
\end{itemize}

\section{The Arthur-Packard-Rogers Model}
\label{APR}

The idea of cellular games was first developed by Norman Packard
and Brian Arthur at the Santa Fe Institute \cite{packard}; and
first written up by K. C. Rogers, in a Master's thesis at the
University of Illinois under the
direction of Dr. Packard \cite{rogers}. In this model, cells
arranged in a ring play a game, such as the well-known Prisoner's
Dilemma, with each of their nearest neighbors. They play for a
fixed number of rounds. At the end of these rounds, or of a
generation, strategies may change. Successful strategies are most
likely to spread and persist. The Prisoner's
Dilemma is discussed in \cite{poundstone}, \cite{axelrod} and Appendix
\ref{appris}.

For details of this model, see Appendix \ref{Rogers}. The
terms used are described in Definition \ref{cgame}.

The Arthur-Packard-Rogers model can be
summarized as follows:
Cells, arranged in a one-dimensional structure, play
a game, such as the Prisoner's Dilemma, with their neighbors,
for a predetermined number of rounds. The criteria for
success in each round do not change, and are the same
for each cell. Since the degree of success is based only
on the moves of a cell and those of its $r$ nearest neighbors
on each side, this criterion can be encoded in the form of
a table.

The strategies that govern cell move choices
may be different for each cell, may be deterministic or stochastic,
are based on past move history, and are
stored in the form of a table. Strategies may have depth zero,
one, or more.

At the end of these rounds -- that is, at
the end of a generation -- the probability that a cell
keeps its strategy in the next generation is proportional
to the size of its reward variable, which measures its success in the
game.

\begin{defn}
{\bf Cell death}: A cell is said to die if its strategy is deemed
replaceable; that is, it is thought of as unsuccessful. The replacing
strategy is usually derived from the strategies of other cells.
\end{defn}

Finally, if a cell dies at the end of a generation,
the strategy chosen is some combination of the strategies of
its nearest living neighbors. If it contains elements
of both neighbors, crossover is said to occur.

\begin{defn}
\label{crossover}
{\bf Crossover} is the existence, in a new strategy, of
behavior similar to more than one ``parent'' strategy.
\end{defn}

\begin{defn}
\label{parents}
Those cells whose strategies contribute to
the new strategy of a cell are called its {\bf parents}.
\end{defn}

There may also be a small probability of strategy table mutation.

\begin{defn}
\label{mutation}
A {\bf mutation} is said to occur when, after a strategy table
entry has been chosen from a parent cell, it is arbitrarily changed.
\end{defn}

In computer simulations, this is often done with the aid of a
pseudorandom number generator.

This model is not quite the same as the original one used in \cite{rogers}.
In that construction, strategy
replacement was not governed by locality; that is, parent cells were the
most successful in the ring. Thus, the progenitor of the strategy
of a cell was not particularly likely to be nearby.

In this model, however, parent cells are not necessarily the most
successful cells in the ring. Instead, they are the nearest
living neighbors of a cell. Such a model is more comparable with
living systems, because it bases system evolution more completely on
local properties. It is also more easily generalizable to the infinite
case, in which there is one cell for each integer. And it is only
under such a model that one can see the evolution of zones of different
strategies.

\section{Computer Experiments}
\label{comp1}
The Arthur-Packard-Rogers model has been simulated in computer experiments,
with the aid of a pseudorandom number generator. Cell moves are displayed
onscreen, in a form similar to the display of cellular automaton
states. That is, initial moves, for each generation, are shown in
a line on top of the screen; and moves for each round are shown
below the preceding round. In experiments simulating the
Prisoner's Dilemma, or variations, lighter areas indicate
cooperative moves; dark areas, defecting moves. In particular,
in the games illustrated in the accompanying figures,
all strategies are mixed, or
stochastic. That is, there is always at least a
small probability that a move is made
other than the one called for by the strategy.

The experiment illustrated in Figures 1 through
14 simulates a variation of the
Prisoner's Dilemma, the Stag Hunt. The Stag Hunt is modeled on
the dilemma of a member of a pack of hunting animals, such as
wolves or coyotes. If the whole pack hunts together, they can
bring down a stag, which is the highest reward. If a member defects, it
will be able to get a rabbit alone. If the other animals do not
defect, they will have a smaller chance of bringing down a stag,
but it may still be possible; but it is very unlikely that one
animal can bring down a stag all by itself.  Thus, the highest
expected reward is for mutual cooperation; next highest, for
defecting while the other members of the pack cooperate; next,
for mutual defection, and fourth, for cooperating while the other
members of the pack defect. See \cite{poundstone} for more
information on the Stag Hunt; and Appendix \ref{appendix1} for a
more technical discussion of the experiments.

These computer experiments fully suggest the mathematical
interest of the subject. They reveal thought-provoking
behavior, such as:

\begin{itemize}
\item {\sl Zone growth.} Strategies may not evolve in the same
manner in all areas of the ring. Zones of cooperative, defecting
or other consistent behavior may arise and persist for
generations.

\item {\sl Periodic structures.} Cells may alternate between
cooperation and defection, or waves of cooperation may spread
through some or all zones of the ring.

\item {\sl ``Complexity.''} Move patterns may display a
sophistication reminiscent of living structures, or the patterns
found in ``complex'' cellular automata.

\item {\sl Long transients.} Strategies predominant for hundreds
of generations may ultimately disappear, and be replaced by
completely different behavior.

\item {\sl ``Punctuated equilibria.''} Move behavior that appears to
be stable for many generations may, suddenly, change very quickly
-- and then become stable again, for a long time.
\end{itemize}

Note that cellular games cannot be construed to represent any
particular living systems, social or biological. For one
thing, their behavior changes very easily as parameters are
modified; it is difficult to tell which features are
essential, or appropriate to any particular model. However,
the existence of the above characteristics suggests that cellular
games are evocative of biological evolution. It seems
possible that the two will turn out to have some features
in common.

\section{The Zero-Depth Model}
\label{ZDM}

Now, these experiments well suggest the richness of behavior
cellular games offer. The sophistication of patterns displayed
provides ample justification for further study of this paradigm.
But the Arthur-Packard-Rogers model does not lend itself well
to mathematical analysis. Its computer implementation is
lengthy and contains many modifiable parameters. It is
difficult to decide if any behavior exhibited is general, or just
an artifact of the specific algorithms used.

To facilitate mathematical discussion of cellular game behavior,
it is hence appropriate to simplify the model. Extensive study
has been performed on such a model, exhibiting the following
simplifications:

\begin {itemize}
\item
{\sl Elimination of crossover.} The Arthur-Pack\-ard-Ro\-gers mo\-del
al\-lows  cross\-over. (Definition \ref {crossover}.)

In the simplified model, crossover is eliminated, and
each new strategy is an exact copy of one that already exists.
A rationale for this simplification, in terms of
living systems, is that one is considering the evolution of a
specific gene, which spreads on an either-or basis. However,
a particular gene may be significant only in the context of
other factors. It may thus not be appropriate to consider this gene
on its own. Note that computer experiments using genetic algorithms
reinforce the significance of crossover (see \cite{levy}).

\item
{\sl Elimination of mutation.} Another simplification is the
elimination of mutation (Definition \ref {mutation}).
That is, after the initial round, any
strategy is new for a specific cell only, and is a copy of the
strategy used by an existing cell. Particularly without crossover,
this elimination is actually
likely to change the long-term behavior of the system. For
example, suppose strategy A is successful against all other
strategies, including itself. If a ring of cells is originally
free of strategy A, but mutation is allowed, strategy A will
eventually take over the ring. If there is no mutation, the ring
will stay free of it. However, the behavior of a cellular game
that allows mutation may best be understood in terms of, and in
comparison to, the behavior of the simpler system.

\item {\sl One round per generation.} That is, cell strategy
may change after each round of play.

\item {\sl Elimination of mixed strategies.} Strategies
are deterministic, not stochastic.

\item
{\sl Elimination of depth.} The final simplification is the
elimination of depth. That is, all strategies are executed
without regard to past moves. Since there are no mixed
strategies, the strategy, then, just becomes
``do move $m$,'' and the move variable can thus be eliminated
from the description of the game.
\end{itemize}

The question of how depth and round restrictions
affect cellular game behavior is a subject for future research; however,
these restrictions are not as severe as they seem. From game theory,
we learn that all information about games with extremely sophisticated
strategies can be conveyed in table form; that is, the
``normal'' form of a game. The only restriction is that strategies
must take into account only a finite amount of information; e.g.,
the course of the game, but not anything before or beyond.
As previously discussed, such tables can be used as the score
table for a cellular game; in particular, for a zero-depth, one
round per generation cellular game.

As a matter of fact, cellular games of many rounds
per generation, and with high-depth strategies, can be rewritten
as zero-depth one round games -- if all strategies take
into account the current generation only.

Note that the Arthur-Packard-Rogers
model, discussed above, does take into account moves in the
previous generation. However, it could easily be modified
not to do so, by providing table entries to use when
there is limited information about previous rounds. For example,
there could be an entry for the move used if nothing is known
about previous moves.

\begin{thm}
Let $G$ be a cellular game of radius $r$, with $R$ rounds per generation,
and strategies of depth $d$ -- except that all strategies take into
account only moves in the current generation. Then the action
of $G$ can be exactly simulated by a cellular game $G'$ of zero
depth and one round per generation.
\end{thm}

{\sl Proof.} It suffices to show that for every such game $G$
there is a zero-depth, one round cellular game $G'$, and a mapping
$f$ from strategies in $G$ to strategies in $G'$, such that life
probabilities correspond. Actions made after cell survival is
decided can be the same in each case.

That is, suppose there are
two rings of $k$ cells each, $1 \le k \le \infty$. Let the first ring
run $G$ in generation $g$, and let each cell $c$ use strategy $S_c$.
Let the second ring run $G'$ in that generation, and let each cell $c'$
use strategy $f(S_c)$. Then the probability, at the beginning of
$g$, that $c$ survives into the next generation should be the
same as the probability that $c'$ does.

To show that such an $f$ can be constructed, it suffices to show
that the probability that, under $G$, at the beginning of a generation, that a
cell will live through to the next generation is entirely
dependent on its strategy, and those of its $(R-1) r$ nearest
neighbors on each side. For if this is true, a table can be
constructed, giving the life probability for cell $c$ if it
and its neighbors follow strategies $S_{c - (R-1) r}, \ldots,$ $S_c,
\ldots, S_{c + (R-1) r}$; and this table can be used to
create a zero-depth, one round cellular game with corresponding
life probabilities.

Now life probabilities in $G$, at the end of a generation, are
entirely dependent on the move histories of that generation. Therefore,
to show such strategy dependence, it is only necessary to show
that the probability, at the beginning of $g$, that cell $c$ will
make move $m$ in generation $q$, is entirely dependent on the
strategies of $c$ and those of its $(q-1) r$ neighbors on each side.

This is trivially true in the first round of a generation.
Since a cell has no information about past moves, the probability
it makes move $m$ is entirely dependent on its own strategy.

Now, suppose this is true for the first $q-1$ rounds. In round
$q$, the probability a cell makes move $m$ is entirely
dependent on its strategy, and the moves made by it and its
$r$ neighbors on each side, in preceding rounds of this generation.
Therefore, by the induction hypothesis, this probability
at the beginning of a generation is entirely dependent
on the strategies of the $(q-2) r$ neighbors of {\sl these}
cells -- cells $c - (q-1) r$ through $c + (q-1) r$.
\rule{2mm}{3mm}

We are thus left with the following model, in which, associated
with each cell $c$, in each generation $g$, are:

\begin{itemize}
\item A move/strategy variable $m_{c,g}$ from some finite alphabet
$\Sigma$ of $k$ characters.

\item A binary-valued life variable $L_{c,g}$. This variable can
be set to either living, or not living.

\end{itemize}

In each generation, cell strategies change, as follows:

\begin{itemize}
\item
The probability that the life variable of a cell is set to 1, so that
it ``lives'' into the next generation, is determined
by a universal and unchanging game matrix $G$. That probability
is based on the move/strategies of a cell and those of its $r$ nearest
neighbors on each side, in that generation.

\item A live cell keeps its strategy in the next generation.

\item A cell that does not live is given a
new strategy in the next generation. This strategy is either that
of its living nearest neighbor to the left, or to the right, with
a 50\% probability of each. If there are no living neighbors
to either side, all possible strategies are equally likely.
\end{itemize}

Note that, in this model, exactly two decisions are
made in a generation; first, decisions about cell
life or death; and second, decisions, for dead
cells, of color in the next generation.

This model lends itself easily to computer simulation, with the
different strategies represented by different colors. Thus, in
descriptions of this model, ``move,'' ``strategy,'' and ``color''
are equivalent. Such a simulation is presented
at the end of this paper, in Figure \ref{cl}. In this simulation,
a cell has probability
$0.27$ of living if it is the same color as both of its neighbors
and $0.53$ otherwise. Due to the shapes of the space-time zones
produced, this process is called the Cloud Process. The Cloud
Process is an example of a join/mix cellular game, as discussed
in Section \ref {examples}.

We now discuss a theorem pertinent to this model; that is, a
simple characterization of identity games. An identity game is a
game in which, outside of certain pathological cases, no cell can
change color. To avoid complications arising from these
cases, the identity game is formally defined as follows:

\begin{defn}
The {\bf identity game} is a game in which, under at least some
circumstances, cells have positive probability of living; and in
which no cell can change strategy, unless there are no living
cells either to the left or right of it.
\end{defn}

The characterization is:

\begin{thm}
Under the zero-depth model, a cellular game is the identity game
if and only if the probability that a cell stays alive, if its
strategy is different from at least one of its neighbors, is 1.
\end {thm}

{\sl Proof.} Suppose a $G$ is a zero-depth cellular game of radius $r$,
with life probabilities fitting the above description.
Suppose a cell has living neighbors on each side. Then either:

\begin {enumerate}
\item A cell is not the same color/strategy as both of its neighbors.
Then it will stay alive.

\item A cell $c$ is the same color as both of its neighbors, but
has neighbors on both sides of different colors, the nearest ones
being cells $c - r_1$ on the left and $c+r_2$ on the right.
Then cells $c-r_1+1$ and $c+r_2-1$ are alive.
Therefore, if $c$ dies, the left parent of $c$ will be cell $c-r_1+1$,
or a cell
closer to $c$; and the right parent of $c$ will be cell $c+r_2-1$, or a
cell closer to $c$. Thus if $c$ dies, both parents will be the same color
as $c$.
\end{enumerate}

On the other hand, suppose $G$ is such that there is positive
probability a cell $c_1$ of color $a$, next to a cell $c_2$ of color
$b$, may not live. Let there be a configuration of
cells giving positive life probability to the center cell.
Thus, since life probabilities
are determined locally, it is possible that there may be living
cells on either side of $c_1$. Let $c_1$ die, and let it
have living neighbors on each side. If either of these neighbors
is not the same color as $c_1$, then $c_1$ may change color; if both are,
$c_2$ will change color.
\rule {2mm}{3mm}

Finally, if cellular games, as described above, are intended to
model living systems, two questions arise. First, why is a new
strategy a symmetric function of the strategies of both parents,
instead of, for example, being more influenced by the strategy of
the nearest parent?

One answer is that this process is intended to model sexual
reproduction, in which a gene has an equal possibility of coming
from each parent. Another is that if there is {\sl positive }
probability that each gene comes from each parent, the model may
actually not behave very differently. Future research may settle
this question.

The second question is, why nonlocality? That is, why not say
that if a cell has no living neighbors near enough, it just stays
dead in the succeeding generation? In this case, comparison with
living ecosystems does suggest that locality is more appropriate,
but with a very large radius. That is, suppose there is a large
die-off of organisms in one particular area. Then organisms from
surrounding areas will rush in very fast, to fill the vacant area
-- but they cannot rush in infinitely far in one generation. Once
again, future research may settle whether the simplified
assumption, that is, nonlocality, actually creates different
long-term behavior.

\section {Ring and Torus Viability}
\label{rvsec}
The following theorem describes move behavior which results in
optimal cell viability, for a whole ring of cells. It applies to
all cellular games with a local life probability matrix; that is,
all games in which the probability a cell ``lives'' into the next
generation is determined by its moves, and those of its neighbors
less than a given number $r$ of units away. It thus applies to
the Arthur-Packard-Rogers model. However, it is here described in
terms of the one-round model given in the previous chapter.

\begin{defn}
\label{ringviability}
The {\bf ring viability} of a finite ring of cells $C$
running a one-round game $G$, in generation $g$, is the average
life probability of these cells in that generation after moves
are made, but before the life variables of the cells are actually set.
\end{defn}

Since $C$ has finitely many cells, whose moves are from a
specific finite alphabet, there is some combination of moves
which will maximize this viability. For example, in a one-round
version of the Stag Hunt game, ring viability will be maximized
if all cells cooperate; and, in some versions of the Prisoner's
Dilemma, ring viability will be maximized if cells alternate
between cooperation and defection.

The result obtained is that this optimal arrangement is periodic.
The following lemma is used in proving this:

\newtheorem{lem} [defn] {Lemma}
\begin{lem}
\label{l1}
Let $G$ be a one-round cellular game of radius $r$, in which
there are $k$ possible moves from some finite alphabet $\Sigma$.
Let $t$ be any string in $\Sigma^*$. Let $L(t)$ be the
average life probability of all cells in a ring of $\vert t \vert$
cells, such that the move of the $i$th cell is the $i$th character of $t$.
Then, if $b$, $w_1$, $w_2$ are strings in $\Sigma^*$,
$\vert b \vert \ge 2r$, then we have

\begin{equation}
\label{ringvia}
L(bw_1bw_2) = { {L(bw_1) + L(bw_2)} \over 2}
\end{equation}
\end{lem}

{\sl Proof.} Consider a ring of cells making consecutively the moves in
$bw_1bw_2$.
Cells making moves from $w_1$ are more than $r$ units away from
cells making moves from $w_2$. Therefore, these cells cannot
influence each other's life probabilities. In the same way, $b$ is
large enough so the life probabilities of cells making moves in either
copy of $b$ can be influenced by cells making moves in $w_1$, or in
$w_2$, but not by both. Therefore the average life probability of all cells
is the same as if they were considered to be in two different rings.
\rule {2mm}{3mm}

The main result follows:

\begin{thm}
\label{rv}
Let $G$ be a one-round cellular game as above. Then
there is some $m > 0$ and some sequence $t$ of $m$ moves,
such that rings
of $nm$ cells, in which the moves of $t$ are repeated $n$ times,
have the maximum ring viability, under $G$,
for finite rings of any size.
\end{thm}

{\sl Proof.}
There are only a finite number of strings in $\Sigma^*$ that either
contain no more than $4r$ letters, or,
when circularly arranged, no duplicate, nonoverlapping $2r$-tuples. Let such
strings be called ``good''; and let $t$ be any ``good'' string
that maximizes $L(t)$. We wish to show that

\begin{equation}
\label{v1}
L(t) = \max_{s \in \Sigma^*} L(s)
\end{equation}

\noindent because, then, rings repeating the moves of $t$ one or more times
would have maximal viability.

Now, this is trivially true for $s$ such that $\vert s \vert \le 2r$,
because all such $s$ are good. Suppose it is true for all $s$ such that
$\vert s \vert < n$. We wish to show that it is true for $s$, such that
$\vert s \vert = n$.

If $s$ is good, this is trivially true. Suppose $s$ is not good. Then
we have $s
= w_1 b w_2 b$, $\vert w_1 \vert$, $\vert w_2 \vert \ge 0$, $\vert b
\vert = 2r$. Lemma \ref{l1} shows that
\begin{equation}
\label{ringvia2}
L(w_1 b w_2 b) = { {L(w_1 b) + L(w_2 b)} \over 2}
\end{equation}

And, by our induction hypothesis, we know that $L(w_1 b) \le L(t)$
and $L(w_2 b) \le L(t)$.
\rule {2mm}{3mm}

A corollary to this theorem is concerned with asymptotic
viability of doubly infinite arrays of cells.

\begin{defn}
Let $l(c)$ be the life probability of a cell $c$, given its move
and those of its $r$ neighbors on each side.
\end{defn}

\begin{defn}
Let the {\bf asymptotic viability} $L(I)$, of a doubly infinite array of
cells $I$, be measured as follows:

\begin{equation}
\label{v2}
L(I) = \limsup_{n \rightarrow \infty} { {\sum_{i= -n}^n l(I_i)} \over
{2n+1} }
\end{equation}
\end{defn}

\newtheorem{cor} [defn] {Corollary}
\begin{cor}
Let $I$ be a doubly infinite array of cells.
Then if $t$ is that finite string that maximizes L(t), $L(I)$
cannot be greater than $L(t)$.
\end{cor}

{\sl Proof.}
Consider what life probability cells $n$ through $-n$ would have
if they were arranged in a ring, instead of part of a doubly
infinite lattice. The only cells that might have different life
probability are cells $-n$ through $-n+r-1$ and $n$ through $n-
r+1$. And as $n$ becomes larger, the contribution of these $2r$
cells to ring viability goes to $0$.  \rule {2mm}{3mm}

In the two-dimensional case, however, a result similar to Theorem
\ref{rv} is false. That is, there are two-dimensional cellular
games, for which no finite torus can achieve maximal torus
viability. This is not shown directly, but is a corollary of
results about Wang tiles.

A Wang tile is a square tile with a specific color on each side.
A set of Wang tiles is a finite number of such tiles, along with
rules for which colors can match. For example, a red edge may be
put next to a blue edge, but not a white edge. Such a set is
said to tile the plane, if the entire plane can be covered by
copies of tiles in the set, so that all edge matchings follow the
rules. Robert Berger \cite{berger} showed that there is a
set of Wang tiles that can tile the plane, but permit no
periodic tiling. Raphael Robinson \cite{robinson} subsequently
discovered another, smaller and simpler set of tiles that
does the same thing.

Note that the set of tiles described by Robinson admits an
``almost periodic'' tiling. That is, for any positive integer
$N$, the plane can be covered with these tiles periodically so
that, under the given rules, the proportion of tiles having
unmatching edges is less than ${1 \over N}$.

A two-dimensional cellular game can be made from a $k$-
colored set of Wang tiles as follows: Let a cell be considered a
tile; let there be $k^4$ possible moves, and let these moves be
considered direct products of the colors of the Wang tiles. Let
the life probability of a cell be increased by ${1 \over 4}$ for every
match of a component of its move, with the corresponding
component of the move of its neighbor. For example, ${1 \over 4}$
would be added to the life probability of a cell, if the left
component of its move were compatible to the right component of
the move of its left neighbor.

Suppose a cellular game were made, in this manner, from the set
of tiles described by Robinson. Then no torus could have
viability one, because otherwise there would be a periodic tiling
of the plane using these tiles. However, there are periodic
tilings of the plane for which only an arbitrarily small
proportion of the tiles have unmatching edges. Therefore, since a
periodic tiling of the plane can be considered a tiling of a
torus, there are torus tilings having viability $1 -\epsilon$,
for any $\epsilon > 0$.

The comparison of cellular games and Wang tilings suggests other
possibilities for future research on tilings. For example,
instead of a Wang tiling in which two colors either match or not,
one could consider a tiling in which two colors can partially
match. This would correspond to a cellular game in which more
than two different levels of success were possible.

\section {Strategy Stability}
\label{stabsec}

In the preceding chapter, the concept of ring viability was
discussed. That is, for each cellular game, there is some
periodic combination of moves which maximizes average cell
viability. One might assume that all cellular games would
stabilize with cells exhibiting, or mostly exhibiting, such a
combination of moves. If this assumption were true, questions
about the long-term evolution of cellular games could be
trivially resolved.

However, computer experiments suggest that this is not necessarily
the case. That is, a one-round cellular game is simulated in
which each cell plays the Prisoner's Dilemma with each of its
neighbors. Specifications are:

\begin {itemize}
\item {\sl Radius.} The game is of radius one.

\item {\sl Strategies.} There are two strategies, or colors:
``C,'' cooperate, or white, and ``D,'' defect, or black.

\item {\sl Game Table.} The game life probability table is: $G(CDC) = 1$,
$G(CDD) = G(DDC) = {7 \over 10}$, $G(CCC) = {6 \over 10}$,
$G(DDD) = {4 \over 10}$, $G(CCD) = G(DCC) = {3 \over 10}$, $G(DCD) = 0$.

($G(m_1m_2m_3)$ is the probability of a cell surviving, if the move
of its right neighbor is $m_1$, its own move is $m_2$, and
the move of its left neighbor is $m_3$.)
\end{itemize}

Under these circumstances, maximal ring viability is achieved by
a ring of all-cooperating cells. And yet, computer experiments
simulating this game do {\sl not} show the mostly cooperative
state to be stable. In the simulation depicted in Figure
\ref{fpris}, a small number of defecting cells
are put in the middle of a large ring of cooperators. The
defecting strategy quickly takes over the ring.

The reason for this is that, although defectors do badly
against each other, they do extremely well against cooperators.
Thus, if a small zone of defecting cells is placed in a large
ring of cooperating cells, the area between the leftmost and
rightmost defecting cells tends to expand.

To address such questions more formally, we use the concept of a
domain:

\begin{defn}
\label{domain}
A {\bf domain} is a contiguous row of same-colored cells.
\end{defn}

We would like to examine what happens when a small defecting domain
is placed between two very large cooperating domains. Is the
number of defecting cells in the vicinity of that domain likely
to go up, or down? If it is more likely to go up, we can
reasonably say that cooperative behavior is not stable under
invasion.

Of course, conceivably, each strategy could be
unstable under invasion by the other; that is, there could be a
tendency for large domains of each color to break up into smaller
ones.

Let there be a doubly infinite lattice of cells, running the
Prisoner's Dilemma game described above. Let $B$ be a small, but
greater than one-cell, black domain in this lattice, bordered, in
generation $1$, by two large white domains $W_l$ and $W_r$. Let
$\vert B \vert $ be the number of black cells in $B$ in generation $0$. Let
$\delta B$ equal the number of cells that were white in
generation $1$, and, in generation $2$, have black strategies
descended from the strategies of cells in $B$ -- minus the
number of cells that were in $B$ in generation $1$, and are white
in generation $1$. Thus, $\delta B$ is, roughly, the change in the
number of black
cells in the vicinity of $B$ in the next generation. Finally, let
$c_1$ be the rightmost member of $W_l$, $c_2$ the leftmost member
of $B$, $c_3$ the rightmost member of $B$, and $c_4$ the leftmost
member of $W_r$, in generation $1$.

Now, two terms used in the theorems presented in this
chapter are defined.

\begin{defn}
Let a {\bf black incursion} be a situation in which a black
cell $c$, in $D$, becomes in the next generation the parent of
newly black cells in $W_l$ or $W_r$. If it becomes the parent of
cells in both, let it be regarded as two incursions.
\end{defn}

\begin{defn}
Let the cell $c$, the parent of the newly black cells in
the incursion, be called the {\bf parent} of the incursion.
\end{defn}

\begin{defn}
Let a {\bf white incursion}, and its {\bf parent}, be defined in a similar
manner; that is, a situation in which a white cell becomes the
parent of cells formerly in $B$.
\end{defn}

\begin{defn}
Let a {\bf black incursion possibility} be a situation in which
an incursion into $W_l$ is possible, because $c_1$ has died,
or a situation in which an incursion into $W_r$ is possible,
because $c_4$ has died. Similarly, let a {\bf white incursion
possibility} be a situation in which an incursion into $B$
with parent in $W_l$ is possible, because $c_2$ has died, or
an incursion into $B$
with parent in $W_r$ is possible, because $c_3$ has died.
\end{defn}

We now show that as the size of the bordering white domain
becomes arbitrarily large, the expected size of a
black incursion into that domain (if possible, as explained above),
should approach $5 \over 6$.

\begin{lem} Let $E_n$ be the expected size of a black incursion
into a white domain $W$, given that there is a black incursion
possibility with parent in $B$, and that $\vert W \vert = n$. Then,
under $G$

\begin{equation}
\label{pris1}
\lim_{n \rightarrow \infty} E_n = {5 \over 6}
\end{equation}
\end{lem}

{\sl Proof.} Suppose the nearest cell $w$, in $W$, to $B$ to stay
alive is such that there are $k$ dead cells in $W$ between $w$
and $B$. Then cells in $W$ between $w$ and $B$ have parents of
both colors, and their probability of becoming black is thus ${1
\over 2}$. Now, the probability of there being $k$ such cells to
die, under $G$, given the incursion possibility, is $G(CCC) (1 -
G(CCC))^{k-1} = {3 \over 5} ({2 \over 5})^{k-1}$. That is, each
white cell with two white neighbors has probability $G(CCC) = {3
\over 5}$ of living. Thus

\begin{equation}
\label{pris2}
\lim_{n \rightarrow \infty} E_n =
\lim_{n \rightarrow \infty}
\sum_{k=1}^{n} ({k \over 2}) ({3 \over 5}) ({2 \over 5})^{k-1} =
\sum_{k=1}^{\infty} ({k \over 2}) ({3 \over 5}) ({2 \over 5})^{k-1} =
{5 \over 6}
\end{equation}
\rule {2mm}{3mm}

We also bound the expected size of a white incursion.

\begin{lem} Let $E_m$ be the expected size, under $G$ of a
white incursion into $B$
from a white domain $W$, given that there is a white incursion
possibility with parent in $W$, and that $\vert B \vert = m$.
Then $E_m < {5 \over 4}$.
\end{lem}

{\sl Proof.} Suppose the nearest cell $b$, in $B$ to $W$ to stay
alive is located so that there are $k$ dead cells in $B$ between $b$ and
$B$. Then
cells in $B$ between $b$ and $W$ have parents of both colors, and
their probability of becoming white is thus ${1 \over 2}$.
Now, the probability of there being $k$ such cells to die,
under $G$, given the incursion possibility, is
$G(DDD) (1 - G(DDD))^{k-1} = {2 \over 5} ({3 \over 5})^{k-1}$.
(Since each black cell with two black
neighbors has probability ${2 \over 5}$ of living.) Thus

\begin{equation}
\label{pris22}
E_m = \sum_{k=1}^{m} ({k \over 2}) ({2 \over 5}) ({3 \over 5})^{k-1} <
\sum_{k=1}^{\infty} ({k \over 2}) ({2 \over 5}) ({3 \over 5})^{k-1} =
{5 \over 4}
\end{equation}
\rule {2mm}{3mm}

The main theorem follows:

\begin{thm} Let $B$ be a small black domain on a doubly infinite
lattice, on which the Prisoner's Dilemma game $G$ is run. Let all
variables be as described above. Then, if $\vert B \vert \ge 2$, and $W_l$
and $W_r$ are large enough, the expected value of $\delta B$,
which is roughly the expected change in the number of black cells in the
vicinity of $W$, is positive.
\end{thm}

{\sl Proof.} We examine eight cases, depending on the life
of $c_1$, $c_2$, $c_3$, and $c_4$. Note that $c_1$ and $c_4$ have
probability $G(CCD) = G(DCC) = {3 \over 10}$ of living;
and $c_2$ and $c_4$, have probability $G(CDD) = G(DDC) = {7 \over 10}$.

\begin{enumerate}
\item All four cells live. Then $\delta B = 0$.

\item $c_1$, $c_2$, $c_3$ live, $c_4$ does not (or the reflection
of this case). The probability of this is $2 ({3 \over 10}) ({7
\over 10})^3$. There is one black incursion possibility (with
$c_3$ as the parent), of expected size that approaches ${5 \over
6}$, as the neighboring domain becomes arbitrarily large.

\item $c_1$, $c_2$ live, $c_3$ dies, $c_4$ lives (or the
reflection). The probability of this is $2 ({3 \over 10}) ({7
\over 10}) ({3 \over 10})^2$. There is one white incursion
possibility (with $c_4$ as the parent), of expected size $< {5
\over 4}$.

\item $c_1$, $c_2$ live, $c_3$, $c_4$ die (or the reflection).
The probability of this is $2 ({3 \over 10}) ({7 \over 10}) ({3
\over 10}) ({7 \over 10})$. There is one black incursion
possibility (with $c_2$ or a cell between $c_2$ and $c_3$ as the
parent), of expected asymptotic size ${5 \over 6}$; and there may
be one white incursion possibility (with a cell to the right of
$c_4$ as the parent), of expected size $< {5 \over 4}$.

\item $c_1$ dies, $c_2$ lives, $c_3$ lives, $c_4$ dies. This
case has probability ${7 \over 10}^4$. There are two black
incursion possibilities (with $c_2$ and $c_3$ as the parents),
of expected asymptotic size ${5 \over 6}$ each.

\item $c_1$ dies, $c_2$ lives, $c_3$ dies, $c_4$ lives (or the
reflection). The probability of this is $2 ({7 \over 10})^2 ({3
\over 10})^2$. There is one black incursion possibility (with
parent $c_2$), of expected asymptotic size ${5 \over 6}$; and one white
incursion possibility (with parent $c_4$), of expected size $< {5
\over 4}$.

\item $c_1$ dies, $c_2$ lives, $c_3$ and $c_4$ die (or the
reflection). The probability of this is $2 ({7 \over 10})^2 ({3
\over 10}) ({7 \over 10})$. There is one black incursion
possibility (with parent $c_2$), of asymptotic size ${5 \over
6}$; and there may be one white incursion possibility (with
parent to the right of $c_4$), of expected size $< {5 \over 4}$.

\item $c_2$ and $c_3$ both die. The probability of this is ${3
\over 10}^2$. There may not be a black incursion, if every cell
in $D$ dies. There are at most two white incursion possibilities
of expected size $< {5 \over 4}$ each.
\end{enumerate}

Thus, if $\vert B \vert \ge 2$, and $W_l$ and $W_r$ are large
enough, under all cases the expected value of $\delta B$ must
exceed $2 ({3 \over 10}) ({7 \over 10})^3 ({5 \over 6}) -
2 ({7 \over 10}) ({3 \over 10})^3 ({5 \over 4}) +
2 ({7 \over 10})^2 ({3 \over 10})^2 ({5 \over 6} -  {5 \over 4}) +
({7 \over 10})^4 2 ({5 \over 6}) +
2 ({7 \over 10})^2 ({3 \over 10})^2 ({5 \over 6} -  {5 \over 4}) +
2 ({7 \over 10})^3 ({3 \over 10}) ({5 \over 6} -  {5 \over 4}) -
({3 \over 10})^2 2 ({5 \over 4}) = {841 \over 6000}$.
\rule {2mm}{3mm}

However, it is not always the case that, in a two-strategy
system, the ``dominant'' strategy will prevail. One strategy may
lose against another, but do so well against itself that its use
tends to expand. This happens in zero-depth versions of the
previously discussed Stag Hunt, a game similar to the Prisoner's
Dilemma, except that successful cooperation is more profitable
than exploitation. If computer experiments (Figure \ref {fstag})
simulate this game,
giving a high enough premium for mutual cooperation, then
cooperative behavior does tend to prevail. Specifically, the
game has the same radius and number of moves as the Prisoner's
Dilemma game described above. Its table is: $G(CDC) = {10 \over
16}$, $G(CDD) = G(DDC) = {7 \over 16}$, $G(CCC) = 1$, $G(DDD) =
{4 \over 16}$, $G(CCD) = G(DCC) = {8 \over 16}$, $G(DCD) = 0$.

It is possible, using the same techniques as above, to show that
black domains are unstable in this game.

\begin{thm} Let $W$ be a small white domain on a doubly infinite
lattice, on which the Stag Hunt game as described above is run.
Let $B_l$ and $B_r$ be its neighbors, and $\vert W \vert$ its size in
generation $1$. Let $\delta W$ equal the number of cells
that were black in generation $0$, and which
in generation $1$, have
white strategies descended from the strategies of cells in $W$ --
minus the number of cells that were in $W$ in generation $1$,
and are black in generation $2$. Then, if $\vert W \vert \ge 2$, and $B_l$
and $B_r$ are large enough, the expected value of $\delta W$,
roughly the expected change in the number of white cells in the
vicinity of $W$, is positive.
\end{thm}

{\sl Proof.} The same calculations as described above are carried
out, except that white and black are exchanged, and the
probabilities of the Stag Hunt game are used. The asymptotic
expected size of a white incursion, given the possibility of
such, turns out to be $2$. The expected size of a black
incursion, given the possibility of such, turns out to be
less than or equal to ${1
\over 2}$ (since cells that are white and bordered on both sides
by white neighbors cannot die). The asymptotic expected change in
the number of white cells in the vicinity of $W$ turns out to exceed
${223 \over 256}$. \rule {2mm}{3mm}

Nash equilibria of cellular games have also been analyzed \cite{cowan}.

\begin{defn}
\label{SNE}
In a cellular game context, a {\bf symmetric Nash equilibrium}
(SNE) arises if, when the $r$ nearest neighbors of a cell on each side use
strategy $s$, its best response is also to use $s$.
\end{defn}

For example, in the Stag Hunt game described above, both
unilateral cooperation and defection give rise to such
equilibria. That is, if the neighbors of a cell always cooperate
(defect), a cell is best off cooperating (defecting) too.

As with ring viability, it is easy to assume that Nash equilibria
determine the course of a game; that is, that a strategy giving
rise to a symmetric Nash equilibrium is stable under
invasion by other strategies. However, while
the study of Nash equilibria is a promising avenue
to understanding cellular games, such an automatic assumption is
not necessarily the case. For example, in the Stag Hunt,
unilateral cooperation gives rise to a SNE. However, in some
versions of this game, cooperating domains are unstable.
This is because though isolated defecting cells
don't survive well, they are likely to kill off their
neighbors. Thus, they tend to have more descendants than their
neighbors.

The parameters used in this version of the Stag Hunt are
not exactly the same as above. They are: $G(CDC) =  $ ${16 \over
18}$, $G(CDD)$ $ = G(DDC) = {15 \over 18}$, $G(CCC) = 1$, $G(DDD) =
{14 \over 18}$, $G(CCD)$ $ = G(DCC) = $ ${9 \over 18}$, $G(DCD) = 0$.

Computer experiments simulating this process (Figure \ref {fstag2})
do indeed suggest
that white domains are unstable. This result can also be
proved using the same techniques as above.

\begin{thm} Let $B$ be a small black domain on a doubly infinite
lattice, on which the second Stag Hunt game as described above is run.
Let $W_l$ and $W_r$ be its neighbors, and $\vert B \vert$ its size in
generation $1$. Let $\delta B$ equal the number of cells
that were white in generation $1$, and, in generation $2$, have
black strategies descended from the strategies of cells in $B$ --
minus the number of cells that were in $B$ in generation $1$,
and are white in generation $2$. Then, if $\vert B \vert \ge 2$, and $W_l$
and $W_r$ are large enough, the expected value of $\delta B$,
roughly the expected change in the number of black cells in the
vicinity of $B$, is positive.
\end{thm}

{\sl Proof.} The same calculations as described for the
Prisoner's Di\-lem\-ma case are carried out, except that the
probabilities of the second Stag Hunt game are used. The asymptotic
expected size of a black incursion, given the possibility of
such, turns out to be ${1 \over 2}$, since cells that are white and
bordered on both sides
by white neighbors cannot die. The expected size of a white
incursion, given the possibility of such, turns out to be
less than or equal to ${9
\over 14}$ . The asymptotic expected change in
the number of black cells in the vicinity of $B$ turns out to exceed
${311 \over 1008}$. \rule {2mm}{3mm}

Thus, we see that cellular game behavior is difficult to
anticipate. These systems reflect the richness of living
ecologies, in which a species' survival is
determined by how well the organisms of that species
compete with others, how well they cooperate among themselves,
and how many descendants
they have. No one factor automatically decides the issue.

\chapter {Two Symmetric Strategies}
\label{TSS}

\section{Introduction and Definitions}

Under the zero-depth model described previously, the simplest
case to examine is that of games with only two possible strategies.
Let these strategies be called {\sl black} and {\sl white}; and
let a cell using a black (white) strategy be called a black
(white) cell. We thus have the following model.

Associated with each cell, in each generation, are:

\begin{itemize}
\item A binary-valued move/strategy variable.

\item A binary-valued life variable. This variable can
be set to either living, or not living.

\end{itemize}

In each generation, cell strategies change, as follows:

\begin{itemize}
\item
The probability that the life variable of a cell is set to 1, so that
it ``lives'' into the next generation, is determined
by a universal and unchanging game matrix $G$. That probability
is based on the move/strategies of a cell, and those of its $r$ nearest
neighbors on each side, in that generation.

\item A live cell keeps its strategy in the next generation.

\item A cell that does not live is given a
new strategy for the next generation. This strategy is either that
of its living nearest neighbor to the left, or to the right, with
a 50\% probability of each. If there are no living neighbors
to either side, all possible strategies are equally likely.
\end{itemize}

We wish to understand the long-term behavior of such processes.
For simplicity, we first consider systems with infinitely many
cells. And, to understand their behavior in general, it is
illuminating to first consider their behavior in the following
case, in which the possible
future courses of evolution are countable.

\begin{defn}
Initial conditions in which there are finitely many black
cells are called {\bf finitely describable} initial conditions.
\end{defn}

Note that if there are initially only finitely many black
cells, there will always be only finitely many black cells.
Therefore, it is more appropriate to speak about a game evolving
{\sl under} such conditions, than {\sl from} such conditions.

The following definitions are also used:

A {\bf domain} (Definition \ref{domain} is a contiguous row of
same-colored cells.

\begin{defn}
\label{zone}
Under finitely describable initial conditions, let
the {\bf zone of uncertainty} start with the leftmost black
cell and end with the rightmost one. If there are
no black cells, there is no such zone.
\end{defn}

Now, suppose each cell had
probability $1$ of staying alive, no matter what. Then all dynamics
would be trivial; the system could never change.
We would like to avoid such situations; that is, we
would like to assure that change is always possible.
We would also like to assure that, under initial conditions
as described above, the two domains on either side of the zone of
uncertainty will, almost always, contain infinitely many living cells.
Both ends are achieved by specifying that each cell always has
positive probability of either living or not living.

\begin{defn}
\label{simple}
Let a cellular game as described above; that is, zero depth,
with two strategies,
and the above restrictions on life probabilities, be called a
{\bf simple cellular game}.
\end{defn}

Now, the main problem associated with any stochastic process is
to figure out how it behaves in the long run; not only to
figure out how it may
behave, but how it must behave.

In this chapter, we settle this question, at least partially,
for certain classes of games. That is, we
consider simple cellular games with left/right symmetry,
evolving under finitely describable initial
conditions. We show that for such games, the probability
that the zone of uncertainty will grow arbitrarily
far in one direction only is zero. It must, with
probability $1$, either disappear, or grow forever
in both directions.

How is this proved? First, we use Theorem \ref{mix00}, presented
below, a result which applies both to cellular games and other
stochastic processes. This theorem implies that if a simple
cellular game evolves as above, and if, under any conditions, the
probability this zone will ``glide'' arbitrarily far to the left
is positive, there are initial conditions under which this
probability can be made as high as desired; that is, greater than
$1 - \epsilon$, for any $\epsilon > 0$.

Then, we show that under such initial conditions $I_{\epsilon}$,
with very high probability of the zone of uncertainty
``gliding'' off in one direction, there would have
to be probability
greater than some constant that another glider
will spin off and shoot out in the other direction. This
constant would not depend on the initial conditions,
but only on the game. This part of the proof is
accomplished in the following manner:

First, without loss of generality, we locate $I_{\epsilon}$
so that the rightmost black cell is cell $0$.

Then, we count cases in which the zone of uncertainty
``glides'' arbitrarily far in one direction only. We need
to count cases in such a way that no case is counted
twice. To do this elegantly, we restrict our attention to
particular cases in which this zone moves to the right in
a certain way; that is, those cases in which, just before
this zone moves past cell $0$ for the last time, there
is exactly one nonnegative black cell, at position $r$ or
greater.

In a lemma, it is shown that under any $I_{\epsilon}$, with
$\epsilon$ small enough,
the probability that the ``glider''
will operate in such a way is more than some fixed
proportion $\gamma$ of the probability that a glider will
operate at all. This $\gamma$ is dependent on the game
only, and not on the initial conditions. Thus, the
sum of all such cases must be greater than $\gamma (1 - \epsilon)$.

For each such case, we show there is another case
with probability only a fixed proportion less,
in which another glider goes off in the other direction.
To do this, we use the fact that what happens at the
end of the zone of uncertainty; that is, to some specific, fixed
number of cells, cannot change the probability of
a one-generation history very much.

Thus, we can put a lower bound $\beta$ to the probability that
in generation $g$, the game behaves exactly as in the
case counted above, except that a two, three or four-cell black
domain $D$ is spun off, at a distance from all other
black cells greater than the radius of the game.

We can show that if there is any positive probability of a
glider moving in one direction, there is positive probability
at least $\alpha$ that, if the zone of uncertainty contains
only a domain like $D$:

\begin{enumerate}
\item This zone will act like a glider, moving arbitrarily
far to the right.
\item This zone will, in every generation, contain more than
one black cell.
\end{enumerate}

Note that this $\alpha$ will also apply if the positive
cells are as above, and the negative black cells
$D$ itself acts as a glider, moving to the right
and staying from that point on in the positive area, and that
this glider from that point on continues to contain two or
more cells. Since the negative black cells are themselves acting
as a glider, it can be shown that they will not interfere with
the behavior of cells in the positive area. It is in this part of
the proof that the left/right symmetry comes in; it is used to
show that gliders can move in both directions.

Since this right-traveling glider continues to contain
two or more cells, we are able again to avoid counting cases twice.
That is, each case is assigned to the last generation in which
there is exactly one nonnegative black cell.

Thus, the probability that the domain between the two gliders
will grow arbitrarily large, and the zone of uncertainty will
continue to expand forever in both directions, can be given a
lower bound. It can be shown, for small enough $\epsilon$, to be
greater than $\gamma \beta \alpha (1 - \epsilon)$,
with these constants depending only on $G$. If $\epsilon$
is small enough, this forces a contradiction. In reference to
these two gliders, this main theorem, Theorem \ref {main}, is
called the Double Glider Theorem.

Another kind of initial condition is also
discussed; that is, initial conditions under which
there is a leftmost white cell and a rightmost
black cell. A conjecture is presented which applies
to such conditions.

Processes that are symmetric black/white, as well
as right/left, are discussed. They are separated
into two categories, mixing processes and clumping
processes. This separation is based on their behavior under
standard restricted initial conditions.
The properties of clumping processes are further examined.
In this context, a theorem is used which can be applied
to symmetric random walks in general.

Finally, computer experiments are presented. These
models simulate the evolution of simple cellular
games, with both kinds of symmetry, on a circular lattice.
It is shown how this evolution varies as parameters vary.

The following theorem applies to all discrete-time
Markov chains. It can be used to characterize
cellular game evolution under finitely describable
initial conditions.

\begin{thm}
\label{mix00}
Let $M = \{ X(t), t \in 0, 1, 2, \ldots\}$ be a discrete-time
Markov chain. Let a finite history be a list of possible values for
$X(i)$, $0 \le i \le n$, for some $0 \le n < \infty$.
Let $H$ be any collection of infinite histories, which
can be expressed as a countable Boolean combination of
finite histories. Furthermore, let no finite part of
any history in $H$ determine membership in $H$.
Let the probability
of $H$, under any initial conditions $X(0) = x$, be positive. Then,
for any $\epsilon > 0$, there are initial conditions $I_\epsilon$
such that there is probability $1 - \epsilon$ the infinite history
of this process (that is, the values of $X(0), X(1), \ldots, X(n),
\ldots$) will be in $H$.
\end{thm}

{\sl Proof.}
Let all possible finite histories of $M$, given $X(0) = x$, be
placed in correspondence with open intervals in $(0,1)$ as follows:

\begin{enumerate}
\item $If P_{xi} > 0$, let the event that $X(1) =i$ correspond
to the open interval
$(\sum_{j < i} P_{xj}, \sum_{j < i} P_{xj} + P_{xi})$.

\item Suppose $X(n) = s$ in generation $n$, $n \ge 1$.
Let the interval $(a,b)$ correspond to the
values of  $X(0) \ldots X(n)$. Then, if $P_{si} > 0$, let the event
that $X(n+1) = i$ in this generation correspond to the open
interval $(a +  \sum_{j < i} P_{sj} (b-a),a +
\sum_{j < i} P_{sj} + P_{si} (b-a) )$.
\end{enumerate}

Similarly, let countable Boolean combinations of finite histories
correspond to countable Boolean combinations of history intervals.
Note that under this relationship, the probability of any finite
history equals the length of the interval; and
the probability of any countable boolean combination of finite
histories $H$ equals the Lesbegue measure of the corresponding measurable
subset of $(0,1)$. Thus,
if $H$ has positive probability, it corresponds to a real subset $S$ of
(0,1) of positive measure.

By a theorem of real analysis \cite {rudin}, if $S \cap (0,1)$
has positive measure, there is some point $p$ contained in
$(0,1)$ such that

\begin{equation}
\label{analysis}
\lim_{\epsilon \rightarrow 0} { {\mu(S \cap (p - \epsilon, p + \epsilon))
}\over {2 \epsilon} } = 1
\end{equation}

\noindent By the construction, there is a history interval contained
in every interval on the unit line. Hence, for every $\epsilon >
0$, there is a history interval $I$, corresponding to a finite
$n$-step history $h$ in which $X(n) = s$, such that ${ {\mu(I
\cap S)} \over {\mu(I)} } \ge 1 - \epsilon$. By the construction,
then, the probability that the future history of $M$ will be in
$H$, given $h$, exceeds $1 - \epsilon$. By the Markov property of
$M$, and the fact that the finite history $h$ does not determine
membership in $H$, the probability of this, given $X(0) = s$,
must also exceed $1 - \epsilon$. \rule {2mm}{3mm}

Note that for this theorem to apply, $H$ must be such that
no finite history determines membership in $H$. For example,
$H$ cannot be all histories such that $X(2) = 1$.
On the other hand, $H$ could be all histories
such that $X(n) = 1$ for infinitely many $n$.

\begin{cor}
\label{mix0}
Let $G$ be any simple cellular game. Let it evolve under finitely
describable initial conditions. Let $H$ be any countable Boolean
combination of finite game histories. Let the probability
of $H$, under any initial conditions, be positive. Then,
for any $\epsilon > 0$, there are finite initial conditions such that
there is probability $1 - \epsilon$ the infinite history of this
game will be in $H$.
\end{cor}
{\sl Proof.} Let the state $X(g)$ of $G$ in generation $g$
be a list of black cells at the beginning of that generation.
Thus, the states of $G$ can be matched with
the positive integers.
The evolution of $G$ can be considered a Markov chain,
since the probability of entering any state is dependent on
conditions in the previous generation only.
\rule {2mm}{3mm}

\section {The Double Glider Theorem}
\label{doubleglider}

The Double Glider Theorem applies to all simple cellular
games with left/\-right symmetry. It shows that if such a
game evolves under finitely describable initial conditions,
the probability that the zone of uncertainty will expand arbitrarily
far in one direction only is zero. That is, the zone of uncertainty cannot
``glide'' forever to the left, or right. It is
shown that if such a glider could evolve, as it
progressed it could throw off a reflected glider, moving in
the opposite direction; and that if both such actions had positive
probability, there would be a contradiction.

A new definition is used in the implementation of this proof.

\begin{defn}
Let the {\bf effective zone of uncertainty} consist, in
each generation, of cells in the following categories:

\begin{enumerate}
\item Cells in the zone of uncertainty.
\item Cells beyond the zone of uncertainty that have a black
cell as one of their nearest living neighbors.
\end{enumerate}
\end{defn}

That is, cells beyond the zone of uncertainty that can
become either black or white are also in this zone.
The extent of this zone in generation $g$ is dependent not
only on cell colors at the beginning of that generation, but
on life/death decisions made during that generation.

Thus, the evolution of a simple cellular game, under finite
initial conditions, can be considered to occur in each
generation as follows:
First, life/death decisions are made about cells within
the zone of uncertainty. Then, if the leftmost living
cell in the zone of uncertainty is black, life/death decisions are
made about cells to the left of this zone. These decisions start with
the cell on its border, and proceed left until
one lives. Then, if the rightmost living cell in the
zone of uncertainty is black, decisions are made in the
same way about cells to the right of this zone. Finally,
black/white decisions are made. There are no other decisions
that can affect the course of this game.

The concept of effective zone of uncertainty can be extended
to apply to cells on each side of a domain.

\begin{defn}
Let the {\bf left effective zone of uncertainty} $D_l$ of
a white domain $D$ consist of:

\begin{enumerate}
\item Those cells in the effective zone of uncertainty to
the left of $D$.
\item Those dead cells in $D$ whose nearest living neighbor
to the left is black (and thus to the left of $D$).
\end{enumerate}

Let the {\bf right effective zone of uncertainty} $D_r$ be
defined similarly.
\end{defn}

Thus, cells that are in $D$, and not in either $D_l$ or
$D_r$, must stay white. We now show that if these two
effective zones stay separated far enough, they
cannot affect each other.

\begin{thm}
\label{newest}
Let $G$ be a simple cellular game of radius $r$, operating
under finite initial conditions.
Let $D$ be a white domain under $G$. In generation $g$,
let $D$ include at least cells $0$ through $r$.
Furthermore, let all
cells in $D_l$ be to the left of cell $0$ and all cells
in $D_r$ be to the right of cell $r$. Then the life/death
probability of any cell in $D_l$ ($D_r$) will not have
been influenced by that of any cell in $D_r$ ($D_l$).
Also, black/white decisions for all cells in $D_l$ ($D_r$)
will be exactly the same as if $D_r$ ($D_l$) did not exist;
that is, if $D_l$ ($D_r$) comprised the entire effective zone
of uncertainty.
\end{thm}

{\sl Proof.} The first statement is true because $\vert D \vert
\ge r + 1$. The second statement is true because if the
effective zone is as thus stated, each cell in $D_l$ ($D_r$)
must have at least one parent in $D_l$ ($D_r$), and
no dead cell can have parents
from both $D_l$ and $D_r$ unless both parents are white.
\rule {2mm}{3mm}

The following lemmas characterize the expansion of
the zone of uncertainty.

\begin{lem}
\label{l11}
Let $G$ be a simple cellular game with left/right symmetry. Let $R(g)$
be the position
of the right border of the zone of uncertainty in generation $g$,
if it exists.
Let $\alpha_1$ be the smallest probability that any
cell stays alive, and $\alpha_2$ the largest.
Then, for any $n$, there is always probability at least
${1 \over 2} (\alpha_1)^2 (1 - \alpha_2)^{n+1}$
that $R(g+2) - R(g) > n$;
and probability at least ${1 \over 2}^{n+2} \alpha_1^4 (1 - \alpha_2)^{n+2}$
that $R(g) - R(g+2) > n$.
\end{lem}

{\sl Proof.}
Without loss of generality, assume $R(g) = 0$; that is, assume that
cell $0$ is black and there are no black cells to the right
of it. Thus, there is
probability at least $\alpha_1 (1 - \alpha_2)^{n+1}$ that,
in generation $g$, cell $0$ lives, and all cells between it and cell $n+2$
do not. Given these events, there is probability at least ${1 \over 2}$
that cell $n+1$ becomes black in that generation. Given these
events, in generation $g+1$ there is probability at least
$\alpha_1$ that
cell $n+1$ lives, thus staying black into the next generation.
Thus there is probability at least
${1 \over 2} (\alpha_1)^2 (1 - \alpha_2)^{n+1}$ that
$R(g+2) - R(g) > n$.

Now, suppose cell $-n-2$ is black. Then there is probability at least
$(\alpha_1)^2 (1 - \alpha_2)^{n+2}$ that, in generation $g$, cell $1$,
which is white, lives, cell $-n-2$ lives,
and all cells between those two do not. Given
these events, there is probability ${1 \over 2}^{n+2}$ that cells
$0$ through $-n$ become white, and cell $-n-1$ black, in
that generation. Given these events, in generation $g+1$ there
is probability at least $(\alpha_1)^2$ that cells $-n$ and
$-n-1$ both live. This will ensure that at the beginning
of generation $g+2$, the zone of uncertainty will still
exist and have the desired border.

On the other hand, suppose cell $-n-2$ is white. Then
there is probability at least $\alpha_1^2 (1 - \alpha_2)^{n+1}$ that,
in generation $g$, cell $0$, which is black, lives, cell $n-2$
lives, and all cells between these two do not. Given
these events, there is probability ${1 \over 2}^{n+1}$ that
cell $-n-1$ becomes black, and cells $-n$ through $-1$
become white in that generation.
Given these events, in generation $g+1$ there is probability
at least $\alpha_1^2 (1-\alpha_2)$ that cells $-n-1$ and $-n$
live and cell $0$ dies. As before, this will ensure that
at the beginning of generation $g+2$, the zone of uncertainty
will still exist and have the desired border. Thus, there is
probability at least ${1 \over 2}^{n+2} \alpha_1^4 (1 - \alpha_2)^{n+2}$
that $R(g) - R(g+2) > n$.
\rule {2mm}{3mm}

Similar results, of course, apply to $L(g)$.

\begin{lem}
\label{l2}
Suppose the zone of uncertainty moves arbitrarily far to the left
only. Then the probability that its right border will not recede
arbitrarily far to the left (that is, that it will stay within
some bounded interval) is $0$. Furthermore, the probability that
the right effective border will not also recede arbitrarily
far to the left is $0$.
\end{lem}

{\sl Proof.}
Let $\alpha_1$ be the smallest probability that any
cell stays alive, and $\alpha_2$ the largest.
Let $R(g)$ be as above. By Lemma \ref{l11}, if
$-k < R(g) < k$ there is probability at least
${1 \over 2} (\alpha_1)^2 (1 - \alpha_2)^{k+n+1}$
that $R(g+2) > n$. Thus, if $-k < R(g) < k$ for infinitely many $g$,
then $R(g)$ will almost always, infinitely many times, be greater
than any $n$.

Let $R'(g)$ be the position of the right border of the
effective zone of uncertainty in generation $G$. (Again, let $R'(g)$ be
defined only if this zone exists.) Each time $R'(g) > -k$, either
$R(g) > -k$, or cell $-k$ has $50\%$ probability of becoming black.
If this cell does become black, $R(g+1)$ will exceed $-k$.
Thus if $R'(g)$ exceeds $-k$ infinitely many times, $R(g)$ will,
with probability $1$, exceed $-k$ infinitely many times too.
\rule {2mm}{3mm}

As above, similar results, apply to the left border of the
zone of uncertainty.

Some concepts are now presented for subsequent use.

Let a cell history for generations $g$ up to $h$ consist of:

\begin{enumerate}
\item The system state (that is, the positions of all black cells)
at the beginning of generation $g$.
\item All meaningful life decisions made in generations $g$ through
$h-1$; that is, all life decisions made within the zone of uncertainty,
and for those cells outside it whose nearest living neighbor on one side
is black.
\item All color decisions made where color is in doubt; that is,
for cells that die and have nearest living neighbors of different
colors on each side.
\end{enumerate}

Let $H(g,h)$ refer to a cell history as described above. Note
that this description only refers to life decisions made within
the effective zone of uncertainty. Thus, the probability of
any history is affected only by such decisions.

Let the following function be defined for any cell history $h =
H(1,g)$ that starts at generation $1$. Let $F_1(h) = 1$ if, under
$h$, in generation $g$ there is exactly one nonnegative black cell,
at position $r$ or greater. Let $F_1(h) = 0$ otherwise.

Similarly, let $F_2$ and $F_3$ be defined for one-generation
cell histories $h = h(g,g+1)$. Let $F_2(h) = 1$, if in generation
$g$ there exactly one nonnegative black cell, in position
$r$ or greater (that is, if $F_1$ would be $1$ for the
previous history), and, under $h$, in generation $g+1$ there are none;
and let $F_2(h) = 0$ otherwise. Let $F_3(h) = 1$ if
in generation $g+1$ there are two, three or four black nonnegative cells,
both next to each other, and both in positions $r$ or greater.
Let $F_3(h) =0$ otherwise.

The following lemmas are used in constructing the main proof.
The next two lemmas, which compare the probabilities of different
$1$-genera\-tion cell histories, both use the same idea:
Changing what happens to only a specific number of
cells is likely to have only a limited effect on the
probability of the history.

\begin{lem}
\label{mix2} Let $G$ be a simple cellular game.
Then for each $1$-generation history $h$ such that $F_2(h) = 1$,
there is a different $1$-generation history $h'$ such that
all the following apply.

\begin{enumerate}
\item
$F_3(h') = 1$.
\item
$h$ and $h'$ both start with the same system states.
\item
At the end of generation $g$, given history $h'$, the negative
black cells are exactly the same as those at the end of $g$
given $h$.
\item
For any history (starting at generation $1$) $h_0$, we have

\begin{eqnarray}
P(H(g,g+1) = h' \vert H(1,g) = h_0) \ge \\
\beta P(H(g,g+1) = h \vert H(1,g) = h_0)
\end{eqnarray}

with $\beta$ depending only on $g$.
\end{enumerate}
\end{lem}

{\sl Proof.} Let $h$ be a cell history such that $F_2(h) = 1$. That
is, at the beginning of the generation $g$ in which $h$ occurs,
there is one nonnegative black cell $c$. Under $h$, $c$
must die, because in generation $g+1$ there will no longer
be any more nonnegative black cells. Let $b$ be the
nearest cell to $c$, on the left, that stays alive in
generation $g$.

Let $\alpha_1$ be the smallest probability that any
cell stays alive, and $\alpha_2$ the largest.
By the definition of a simple cellular game, both
these numbers must be greater than $0$.
Let $\alpha_3$ be the minimum of $\alpha_1$,
$\alpha_2$, $1 - \alpha_1$, $1 - \alpha_2$.

Case I: $b$ is black (and thus in a negative-numbered position).
Let cell $d$ be the closest living neighbor of cell $c$ on the right.
Let it die as before, and let cell $d+1$ die. As under $h$,
all dead cells between $b$ and $d$ have a $50\%$ chance of becoming black.
Let their colors be assigned the same; e.g., cells $0$ through $d-1$ will
become white. Let cells $d$ and $d+1$ become black. Let all other
life/death and black/white decisions be as under $h$.

Thus, this new history $h'$ satisfies $F_3(h') = 1$,
it produces the same negative black cells as $h$, and we have

\begin{eqnarray}
P(H(g,g+1) = h' \vert H(1,g) = h_0) \ge \\
{(\alpha_3)^2 \over 2} P(H(g,g+1) = h \vert H(1,g) = h_0)
\end{eqnarray}

Also, $h$ can be reconstructed if $h'$ is known; that is:
\begin{enumerate}
\item Initial conditions are the same for both histories.
\item Under $h'$, the location of cells $d$ and $d+1$ are
known; they are the only nonnegative black cells in generation $g+1$.
\item All life/death and color decisions in the effective zone
of uncertainty are the same, except for cells $d$ and $d+1$.
\item The history of cell $d$, under $h$, is exactly known. It
stays alive and stays white.
\item The life or death of cell $d+1$, under $h$, is not known.
However, under $h$, this cell is not in the effective zone
of uncertainty and decisions about it are not considered part
of the cell history.
\end{enumerate}

Thus, in this case, for each different $h$ there is a different
$h'$ satisfying the conditions of this lemma.

Case II: Cell $b$ is white, and one cell to the left of $c$.
Under $h'$, let cells $c$ and $c+3$ live. Let cells $c+1$
and $c+2$ die. Since $c$ is their right parent, they
can become black in the next generation; let them do so.

Let all other cells live or die, and change color,
as under $h$. Note that cell $c$ cannot become a parent
of cells to the left, since it is bordered on the left
by the living cell $b$.

Thus, $F_3(h')$ will be $1$, it will produce the same
negative black cells as $h$, and we have

\begin{eqnarray}
P(H(g,g+1) = h' \vert H(1,g) = h_0) \ge \\
{(\alpha_3)^4 \over 4} P(H(g,g+1) = h \vert H(1,g) = h_0)
\end{eqnarray}

A history $h'$ constructed in this manner cannot be
confused with one created using the first method, since
at the end there are three nonnegative cells rather than
two. Its uniqueness can be shown by methods similar
to those used in the first case.

Case III: Cell $b$ is white and more than one cell to the
left of $c$. Let cells $c-1$ and $c$ live; let
cells $c+1$ through $c+3$ die, and let cell $c+4$ live.
Let all other cells live or die as under $h$.

Now, cell $c-1$ must be white, since cell $c$ is isolated.
Therefore, cells $b+1$ through $c-2$ must, as under
$h$, become white. Let cells $c+1$ through $c+3$
become black. Note that all other cells have the
same color options as under $h$.

Thus, $F_3(h')$ will be $1$, it will produce the same
negative black cells as $h$, and we have

\begin{eqnarray}
P(H(g,g+1) = h' \vert H(1,g) = h_0) \ge \\
{(\alpha_3)^6 \over 8} P(H(g,g+1) = h \vert H(1,g) = h_0)
\end{eqnarray}

This $h'$ cannot be confused with one created using the first
two methods, since at the end there are four nonnegative cells rather than
three or two. Its further uniqueness can also be shown by methods similar
to those used in the first case. Therefore, the conditions
of the theorem are satisfied for all three cases, with
$\beta = { (\alpha_3)^6 \over 8}$.
\rule {2mm}{3mm}

Now, if there is positive probability of a glider --
that is, of the effective zone of uncertainty moving
arbitrarily far in one direction only -- then there is
positive probability that in some generation $g$, this
zone will leave the nonnegative area for the last time.

The following lemma characterizes, for certain initial conditions,
how this can happen. For these conditions, we put a minimum
bound on the probability that, in the generation this zone
leaves the nonnegative area, there is exactly one black cell
-- and this cell is at position $r$ or greater. This bound
depends only on $G$.

The ways this zone can leave the nonnegative area are divided
into four cases. (Actually, three main cases; the last two are
quite similar.) For each of these cases, a different construction is used
to accomplish the proof. As in the preceding lemma, each of
these constructions uses histories that behave similarly
to the ones under consideration, and hence have similar
probabilities of occurrence.

\begin{lem}
\label{mix2a} Let $G$ be a simple cellular game of radius $r$,
operating under finite initial conditions $I$.
Let $\alpha_1$
be the lowest probability, under $G$, that any cell stays alive.
Let there be positive probability that under $I$ the effective zone
of uncertainty moves arbitrarily far to the left; that is,
that some generation $g$ is the last in which the effective zone
of uncertainty contains nonnegative cells. Let $Z_{g} = 1$ if
this is true for generation $g$, and $0$ otherwise. Let
$P(\exists g, Z_{g} = 1)$ exceed $1 - {\alpha_1 \over 2}$.
Let $X_{g} = 1$ if $Z_{g} = 1$, and at the beginning of $g$ there is only one
nonnegative black cell, at position $r$ or greater, and $0$ otherwise.
Then, for some $\gamma$ depending only on $G$

\begin{equation}
P(\exists g, X_{g} = 1) \ge \gamma P(\exists g, Z_{g} = 1)
\end{equation}
\end{lem}

{\sl Proof.}
Let $\alpha_2$ be the highest probability, under $G$, that any
cell stays alive. (By definition, $\alpha_1$, $\alpha_2 > 0$.)
Let $\alpha_3$, again, be the minimum of $\alpha_1$, $\alpha_2$,
$1 - \alpha_1$ and $1 - \alpha_2$.
Let $\alpha_4$ be the life probablity of a black cell
whose $r$ neighbors on each side are also black.
Let $c_{g}$ be the rightmost living cell in the zone of uncertainty,
in generation $g$. Let $D_{g}$ be the rightmost black domain in that zone,
and let $e_{g}$ be the white cell at its left border.

First of all, we know that there is probability at least $\alpha_1$
that in generation $1$, the leftmost black cell lives. Therefore,
there is probability at least $\alpha_1$ that $Z_{1} = 0$. Thus,
if $P(\exists g, Z_{g} = 1) > 1 - {\alpha_1 \over 2}$, we know that
$P(\exists g, g \ge 2, Z_{g} = 1) \ge {\alpha_1 \over 2}$.

Now, suppose there exists a generation $g > 1$ such that $Z_{g} = 1$.
The conditions under which that occurs can be divided into
four cases, as follows:
\begin{enumerate}
\item $c_{g}$, as described above, is black.
\item $c_{g}$ is white, and $c_{g-1}$ is black.
\item $c_{g}$ is white, $c_{g-1}$ is white, and $e_{g-1}$ is alive.
\item $c_{g}$ is white, $c_{g-1}$ is white, and $e_{g-1}$ is not alive.
\end{enumerate}

Let $C_{g}$ be $k$, $1 \le k \le 4$, if case $k$ holds. Thus, there is a
$k$, $1 \le k \le 4$, such that

\begin{equation}
\label{opossum}
P(\exists g, Z_{g} = 1, C_{g} = k) \ge
{1 \over 4} {\alpha_1 \over 2} P(\exists g, Z_{g} = 1)
\end{equation}

Case I. (\ref {opossum}) is true with $k$ set to 1.
In this case, $c_{g}$ is black. Let $d_{g}$ be the living
cell just to the right of the effective zone of uncertainty. For
$Z_{g}$ to be $1$, $d_{g}$ must be at position $1$ or greater.

We wish to show that for each two consecutive $1$-generation histories
$h$, $i$ such that if $H(g,g+1) = h$, $C_{g} = 1$,
there exists a different collection of histories
$h'$, $i'$, such that, for $\kappa$ depending only on $G$, we have

\begin{eqnarray}
\label{a0}
P(H(g,g+1) = h', H(g+1,g+2) \in i') \ge \\
\kappa P(H(g,g+1) = h, H(g+1,g+2) = i)\\
\end{eqnarray}

and

\begin{eqnarray}
\label{a1}
P(X_{g+1} = 1 \vert H(g,g+1) = h',\\
H(g+1,g+2) \in i') =\\
P(Z_{g} = 1 \vert H(g,g+1) = h,\\
H(g+1,g+2) = i)
\end{eqnarray}

Let $h'$ be constructed as follows:

\begin{enumerate}
\item Initial colors are the same as under $h$.
\item Cells $d_{g}$ through $d_{g}+r$ die.
\item Cell $d_{g}+r+1$ lives.
\item All other cells live or die as under $h$. Thus, cells $d_{g}$ through
$d_{g}+r$ are the only ones with different color possibilities than
under $h$; that is, they have a $50\%$ chance of becoming black,
with $c_{g}$ as their parent.
\item Cells $d_{g}$ through $d_{g}+r-1$ become white.
\item Cell $d_{g}+r$ becomes black.
\item All other cells become black or white as under $h$.
\end{enumerate}

At the end of $h'$, we are left with exactly
the same black cells as at the end of $h$, except that cell $d_{g}+r$
is black. And, because of cells added to the zone of uncertainty
under $h'$:

\begin{equation}
\label{a2}
P(H(g,g+1) = h') \ge
{(\alpha_3)^{r+2} \over 2^{r+1}} P(H(g,g+1) = h)
\end{equation}

Also, $h$ can be reconstructed if $h'$ is known; that is:
\begin{enumerate}
\item Initial conditions are the same for both histories.
\item The location of cell $d_{g}+r$ can be recovered. After
the completion of $h'$, it is the right black cell. Hence,
the location of cell $d_{g}$ can be recovered.
\item Under $h$, all life/death and color decisions in the effective zone
of uncertainty, through cell $d_{g}-1$, are the same.
\item Under $h$, cell $d_{g}$ lives, thus bounding the zone of uncertainty.
\end{enumerate}

For $Z_{g}$ to be $1$, in generation $g+1$ the effective
zone of uncertainty must not reach the nonnegative area.
Therefore, $d_{g+1}$ must not be positive.
Let $H(g+1,g+2) = i$ be such a history.
Let $i'$ be constructed as follows, given $i$ and its predecessor $h$:

\begin{enumerate}
\item Let initial colors be the same as under $i$, except
that cell $d_{g}+r$ is black. (The position of $d_{g}$ can
be determined, given $h$.)
\item Let the life of all cells in the effective zone of uncertainty of
$i$ be determined as under $i$.
\item Let cell $d_{g}+r-1$ live. Thus, since the effective
zone of uncertainty of $i$ stays in the negative area, all cells
in this zone will face the same black/white decisions. Also,
cells $d_{g+1}$ through $d_{g}+r-2$ must, if they die, become white.
\item Let cell $d_{g}+r$ die.
\item Let cell $d_{g}+r+1$ live. Thus, cell $d_{g}+r$ will become white.
\item Let all black/white decisions in the effective
zone of uncertainty of $i$ be determined just as under $i$.
\end{enumerate}

In this generation, cells $d_{g+1}$ through $d_{g}+r-2$ can live or
die without affecting the inclusion of a history in $i'$.
Note that the only additional specification for what happens in
$i'$, as opposed to $i$, is the life or death of three particular cells.

Thus, we have

\begin{eqnarray}
\label{a3}
P(H(g+1,g+2) \in i' \vert H(g,g+1) = h') \ge\\
(\alpha_3)^3 P(H(g+1,g+2) = i \vert H(g,g+1) = h)
\end{eqnarray}

Note that $i$ can be recovered, given $i'$, because all
decisions in the effective zone of uncertainty of $i$ are the same.
Also note that conditions after $i'$ are the same
as after $i$. Thus, we have

\begin{eqnarray}
\label{a4}
P(Z_{g+1} = 1 \vert H(g,g+1) = h', H(g+1,g+2) \in i') =\\
P(Z_{g} = 1 \vert H(g,g+1) = h, H(g+1,g+2) = i)
\end{eqnarray}

Since $i'$ starts with exactly one nonnegative cell, at position
$r$ or greater, (\ref{a1}) holds.

Combining (\ref{a2}) and (\ref{a3}), we have (\ref{a0}) holding
with $\kappa = {(\alpha_3)^{r+5} \over 2^{r+1}}$.

Since there is a different $h'$, $i'$ for each different $h$, $i$,
we have

\begin{eqnarray}
P(\exists g, X_{g+1} = 1, C_{g} = 1) \ge\\
\sum_{g,h,i} P(X_{g+1} = 1, C_{g} = 1 \vert\\
H(g,g+1) = h', H(g+1,g+2) \in i')\\
P(H(g,g+1) = h', H(g+1,g+2) \in i') \ge\\
\sum_{g,h,i} P(Z_{g} = 1, C_{g} = 1 \vert\\
H(g,g+1) = h, H(g+1,g+2) = i)\\
\kappa P(H(g,g+1) = h, H(g+1,g+2) = i) =\\
P(\exists g, Z_{g} = 1, C_{g} = 1)
\end{eqnarray}

Thus, by our case hypothesis, we have

\begin{equation}
P(\exists g, X_{g+1} = 1,C_{g} = 1) \ge
{\kappa \over 4} {\alpha_1 \over 2} P(\exists g, Z_{g} = 1)
\end{equation}

Case II. (\ref {opossum}) is true with $k$ set to 2.
In this case, $c_{g}$ is white, and $c_{g-1}$ is black.
Let $d_{g-1}$ be the living
cell just to the right of the zone of uncertainty,
in generation $g-1$. Note that this cell is to the
right of any cells that are black in generation $g$.
Hence, for $Z_{g}$ to be $1$, $d_{g-1}$ must be at position $1$ or greater.

We wish to show that for each three consecutive $1$-generation histories
$k$, $h$, $i$ such that if $H(g,g+1) = h$, $C_{g} = 2$,
there exists a different collection of histories
$k'$, $h'$, $j'$ such that, for $\kappa$ depending only on $G$, we have

\begin{eqnarray}
\label{b0}
P(H(g-1,g) = k',H(g,g+1) \in h',H(g+1,g+2) \in i') \ge \\
\kappa
P(H(g-1,g) = k,H(g,g+1) = h, H(g+1,g+2) \in i)
\end{eqnarray}

and

\begin{eqnarray}
\label{b1}
P(X_{g+1} = 1 \vert H(g-1,g) = k', H(g,g+1) \in h',\\
H(g+1,g+2) \in i') =\\
P(Z_{g} = 1 \vert H(g-1,g) = k, H(g,g+1) = h,\\
H(g+1,g+2) = i)
\end{eqnarray}

Let $k'$ be constructed as follows:

\begin{enumerate}
\item Initial colors are the same as under $k$.
\item Cells $d_{g-1}$ through $d_{g-1}+r$ die.
\item Cell $d_{g-1}+r+1$ lives.
\item All other cells live or die as under $k$. Thus, cells $d_{g-1}$
through $d_{g-1}+r$ are the only ones with different color possibilities
than under $k$; that is, they have a $50\%$ chance of becoming black,
with $c_{g-1}$ as their parent.
\item Cells $d_{g-1}$ through $d_{g-1}+r-1$ become white.
\item Cell $d_{g-1}+r$ becomes black.
\item All other cells become black or white as under $k$.
\end{enumerate}

At the end of $k'$, we are left with exactly
the same black cells as at the end of $k$, except that cell $d_{g-1}+r$
is black. And, because of cells added to the zone of uncertainty
under $k'$, we have

\begin{equation}
\label{b2}
P(H(g-1,g) = k') \ge
{(\alpha_3)^{r+2} \over 2^{r+1}} P(H(g-1,g) = k)
\end{equation}

Also, $k$ can be reconstructed if $k'$ is known; that is:
\begin{enumerate}
\item Initial conditions are the same for both histories.
\item The location of cell $d_{g-1}+r$ can be recovered. After
the completion of $k'$, it is the right black cell. Hence,
the location of cell $d_{g-1}$ can be recovered.
\item Under $k$, all life/death and color decisions in the effective zone
of uncertainty, through cell $d_{g-1}-1$, are the same.
\item Under $k$, cell $d_{g-1}$ lives, thus bounding the zone of uncertainty.
\end{enumerate}

Let $H(g,g+1) = h$ be a history that, together
with its predecessor $k$, satisfies the conditions for $C_{g}$ to be $2$,
and for $Z_{g}$ to possibly be $1$: That is, under $h$, let
the leftmost living cell in the zone of uncertainty be white,
and let this zone leave the nonnegative area.
Let $h'$ be constructed as follows, given $h$ and its predecessor $k$:

\begin{enumerate}
\item Let initial colors be the same as under $h$, except
that cell $d_{g-1}+r$ is black. (The position of $d_{g}$ can
be determined, given $h$.)
\item Let the life of all cells in the effective zone of uncertainty of
$h$ be determined as under $h$.
\item Let cell $d_{g-1}+r-1$ live. Thus, since under $h$ the
the effective zone of uncertainty does not reach this far to
the left, all cells in this zone will face the same black/white decisions.

Note that since under $h$ the left border of the zone of
uncertainty recedes, $e_{g}$ -- that is, the white cell
at the border of this zone -- must be to the right of cell $d_{g-1}+r$.
Also, cells $e_{g}$ through $d_{g-1}+r-2$ must, if they die,
become white.
\item Let cell $d_{g}+r$ live.
\item Let cell $d_{g}+r+1$ live.
\item Let all black/white decisions in the effective
zone of uncertainty of $h$ be determined just as under $h$.
\end{enumerate}

In this generation, cells $e_{g}$ through $d_{g}+r-2$ can live or
die without affecting the inclusion of a history in $h'$.
Note that the only additional specification for what happens in
$h'$, as opposed to $h$, is that three particular cells live.

Thus, we have

\begin{eqnarray}
\label{b3}
P(H(g,g+1) \in h' \vert H(g-1,g) = k') \ge\\
(\alpha_3)^3 P(H(g,g+1) = i \vert H(g-1,g) = k)
\end{eqnarray}

Note that $h$ can be recovered, given $h'$, because all
decisions in the effective zone of uncertainty of $h$ are the same.

For $Z_{g}$ to be $1$, in generation $g+1$ the effective
zone of uncertainty must not reach the nonnegative area.
Therefore, $d_{g+1}$ must not be positive.
Let $H(g+1,g+2) = i$ be such a history.
Let $i'$ be constructed as follows, given $i$ and its predecessors
$h$ and $k$:

\begin{enumerate}
\item Let initial colors be the same as under $i$, except
that cell $d_{g-1}+r$ is black. (The position of $d_{g-1}$ can
be determined, given $k$.)
\item Let the life of all cells in the effective zone of uncertainty of
$i$ be determined as under $i$.
\item Let cell $d_{g-1}+r-1$ live. Thus, since the effective
zone of uncertainty of $i$ stays in the negative area, all cells
in this zone will face the same black/white decisions. Also,
cells $d_{g+1}$ through $d_{g-1}+r-2$ must, if they die, become white.
\item Let cell $d_{g-1}+r$ die.
\item Let cell $d_{g-1}+r+1$ live. Thus, cell $d_{g-1}+r$ will become white.
\item Let all black/white decisions in the effective
zone of uncertainty of $i$ be determined just as under $i$.
\end{enumerate}

In this generation, cells $d_{g+1}$ through $d_{g-1}+r-2$ can live or
die without affecting the inclusion of a history in $i'$.
Note that the only additional specification for $i'$, as opposed
to $i$, is the life or death of three particular cells.

Thus, we have

\begin{eqnarray}
\label{bb3}
P(H(g+1,g+2) \in i' \vert\\
H(g,g+1) \in h', H(g-1,g) = k') \ge\\
(\alpha_3)^3 P(H(g+1,g+2) = i \vert\\
H(g,g+1) = h,H(g-1,g) = k)
\end{eqnarray}

Note that $i$ can be recovered, given $i'$, because all
decisions in the effective zone of uncertainty of $i$ are the same.
Also note that conditions after $i'$ are the same
as after $i$. Thus, we have

\begin{eqnarray}
\label{bb4}
P(Z_{g+1} = 1 \vert\\
H(g-1,g) = k', H(g,g+1) \in h', H(g+1,g+2) \in i') =\\
P(Z_{g} = 1 \vert \\
H(g-1,g) = k, H(g,g+1) = h, H(g+1,g+2) = i)
\end{eqnarray}

Since $i'$ starts with exactly one nonnegative cell, at position
$r$ or greater, (\ref{b1}) holds.

Combining (\ref{b2}), (\ref{b3}),
and (\ref{bb3}), we have (\ref{b0}) holding
with $\kappa = {(\alpha_3)^{r+8} \over 2^{r+1}}$.

Since there is a different $k'$, $h'$, $i'$ for each different $k$,
$h$, $i$, we have

\begin{eqnarray}
P(\exists g, X_{g+1} = 1, C_{g} = 2) \ge\\
\sum_{g,k,h,i} P(X_{g+1} = 1, C_{g} = 2 \vert\\
H(g-1,g) = k',H(g,g+1) \in h', H(g+1,g+2) \in i')\\
P(H(g-1,g) = k', H(g,g+1) \in h', H(g+1,g+2) \in i') \ge\\
\sum_{g,k,h,i} P(Z_{g} = 1, C_{g} = 2 \vert\\
H(g-1,g) = k, H(g,g+1) = h, H(g+1,g+2) = i)\\
\kappa P(H(g-1,g) = k, H(g,g+1) = h, H(g+1,g+2) = i) =\\
P(\exists g, Z_{g} = 1, C_{g} = 2)
\end{eqnarray}

Thus, by our case hypothesis, we have

\begin{equation}
P(\exists g, X_{g+1} = 1,C_{g} = 2) \ge
{\kappa \over 4} {\alpha_1 \over 2} P(\exists g, Z_{g} = 1)
\end{equation}

Case III. (\ref {opossum}) is true with $k$ set to 3.
In this case, $c_{g}$ is white, and $c_{g-1}$ is white.
Let $b_{g-1}$ be the white cell just to the right of $D_{g-1}$. In
this case, $b_{g-1}$ equals $c_{g-1}$.

We wish to show that for each three consecutive $1$-generation histories
$k$, $h$, $i$ such that if $H(g,g+1) = h$, $C_{g} = 3$,
there exists a different collection of histories
$k'$, $h'$, $j'$ such that, for $\kappa$ depending only on $G$, we have

\begin{eqnarray}
\label{c0}
P(H(g-1,g) \in k', H(g,g+1) \in h',H(g+1,g+2) \in i') \ge\\
\kappa
P(H(g-1,g) = k)\\
P(H(g,g+1) = h \vert H(g-1,g) = k)\\
P(H(g+1,g+2) = i \vert\\
H(g-1,g) = k, H(g,g+1) = h)
\end{eqnarray}

and

\begin{eqnarray}
\label{c1}
P(X_{g+1} = 1 \vert H(g-1,g) \in k', H(g,g+1) \in h',\\
H(g+1,g+2) \in i') =\\
P(Z_{g} = 1 \vert H(g-1,g) = k, H(g,g+1) = h,\\
H(g+1,g+2) = i)
\end{eqnarray}

Let $k'$ be constructed as follows:
\begin{enumerate}
\item Initial colors are the same as under $k$.
\item Both the leftmost and rightmost cells in $D$ live
(cells $b_{g-1}+1$ and $e_{g-1}-1$). Thus, all cells in $D$ must become black.
\item Cells $e_{g-1}$ through $e_{g-1}+r$ die.
\item Cell $e_{g-1}+r+1$ lives. Thus, cells $e_{g-1}$ through $e_{g-1}+r$
may become either black or white.
\item All other cells, up to the left border of the zone of uncertainty
of $k$, live or die as under $k$. In specific, $b_{g-1}$ lives, as under
$k$. Thus, all cells to the left of $b_{g-1}$ are faced with the same
black/white decisions as under $k$.
\item Cells $e_{g-1}$ through $e_{g-1}+r-1$ become white, and cell
$e_{g-1}+r$ becomes black. \item All other cells become black or
white as under $k$. \end{enumerate}

Note that all cells in $D_{g-1}$, except for those on each border, can
live or die without affecting the inclusion of a history in $k'$.

At the end of any history in $k'$, we are left with exactly
the same black cells as under $k$, except that all cells in
$D_{g-1}$ are black and cell $e_{g-1}+r$ is black.

Now, consider those cells in the interior of $D_{g-1}$.
Under $k$, they must all die; under
$k'$, their life or death does not matter. On the other
hand, the two cells at the border of $D_{g-1}$ die under $k$,
and live under $k'$. Also, cells $e_{g-1}$ through $e_{g-1}+r$ are
outside the zone of uncertainty under $k$. Under $k'$,
they die, and their colors are specified.

Thus, if $n$ is the maximum of $\vert D_{g-1} \vert - 2r$ and $0$, we have

\begin{equation}
\label{c2}
P(H(g-1,g) \in k') \ge
{(\alpha_3)^{r+3} \over 2^{r+1} (1 - \alpha_4)^{n}} P(H(g-1,g) = k)
\end{equation}

Also, $k$ can be reconstructed if $k'$ is known; that is:
\begin{enumerate}
\item Initial conditions are the same for both histories.
\item Under $k$, all life/death and color decisions in the effective zone
of uncertainty, up to cell $b_{g-1}$, are the same.
\item The location of cell $b_{g-1}$ can be recovered. After
the completion of $k'$, it is the rightmost white cell in the
next-to-rightmost finite white domain.
\item Cell $b_{g-1}$ lives, as under $k'$.
\item The location of cell $e_{g-1}$ can be recovered. After the
completion of $k'$, it is the leftmost cell in the rightmost
finite white domain.
\item Under $k$, cells $b_{g-1}+1$ through $e_{g-1}-1$ die, and become white.
\item Under $k$, cells $e_{g-1}$ and all cells to the right of
it are outside the zone of uncertainty.
\end{enumerate}

Let $H(g,g+1) = h$ be a history that, together
with its predecessor $k$, satisfies the conditions for $C_{g}$ to
be $3$, and for $Z_{g}$ to possibly be $1$. That is, under
both $k$ and $h$, let
the leftmost living cell in the zone of uncertainty be white.
Thus, since the right border of this zone will recede in
generation $g-1$, $D_{g}$ is completely to the left of $D_{g-1}$.
Also, in generation $g$, let this zone leave the nonnegative
area; and let $b_{g-1}$ be alive.

Let $h'$ be constructed as follows, given $h$ and its predecessor $k$:

\begin{enumerate}
\item Let initial colors be the same as under $h$, except
that cell $e_{g-1}+r$, and all cells in $D_{g-1}$, are black.
(The location of $D_{g-1}$, and hence of cells $b_{g-1}$ and
$e_{g-1}$, can be determined given $k$.)
\item Let the life of all cells in the zone of uncertainty of
$h$ be determined as under $h$.
(Note that $D_{g-1}$ is to the right of this zone.)
Furthermore, let cell $e_{g}$ live or die as under $h$.
\item Let the white cells at each border of $D_{g-1}$  -- that is,
cells $b_{g-1}$ and $e_{g-1}$ -- live.
\item Let all cells in $D_{g-1}$ die. Thus, they must all
become white.
\item Let cell $e_{g}+r-1$ live.
\item Let cell $e_{g-1}+r$ live.
\item Let cell $e_{g-1}+r+1$ live.
\end{enumerate}

Note that in generation $g$, cells $e_{g}$ through $e_{g-1}$ --
that is, the cells between the border of the zone of uncertainty
of $h$ and the left border of $D_{g-1}$ -- can
live or die without affecting the inclusion of a history in $h'$.
Also, cells $e_{g-1}+1$ through $e_{g-1}+r-2$, (if $r$ is
large enough for these cells to exist) can live or die
without affecting this inclusion.

At the end of any history in $h'$, we are left with exactly
the same black cells as under $h$, except that cell $e_{g-1}+r$
is black. Now, since $e_{g-1} > e_{g} \ge 0$, we have $e_{g-1}+r) > r$.
And for $Z_{g}$ to be $1$, there must be no nonnegative black
cells at the end of generation $g$. Thus, at the end of $h'$ there
will be only one nonnegative black cell, cell $e_{g-1}+r$.

Now, consider those cells in the interior of $D_{g-1}$.
Under $h'$, they must all die; under
$h$, they are outside the zone of uncertainty. Also, those
cells $r$ or less to the left of $D_{g-1}$ (cells $b_{g-1} - 1$
through $b_{g-1} - r$ may have different life probabilities.
Finally, we have to consider the life probabilities of cells
$e_{g}$, and $e_{g}+r-1$ through $e_{g}+r+1$.

Thus, if $n$ is the maximum of $\vert D_{g-1} \vert - 2r$ and $0$, we have

\begin{equation}
\label{c3}
P(H(g,g+1) \in h') \ge (\alpha_3)^{3r+4} (1 - \alpha_4)^{n}
P(H(g,g+1) = h)
\end{equation}

Also, $h$ can be reconstructed if $h'$ is known. That is,
\begin{enumerate}
\item Under $h'$, $D_{g-1}$ is the second black domain on the left.
\item Under $h$, initial conditions to the right of $D_{g-1}$ are the
same as under $h'$.
\item Under $h$, the zone of uncertainty does not reach $D_{g-1}$.
\item Decisions in the zone of uncertainty of $h$,  and at its border,
both life/death and black/white, are exactly as under $h$.
\end{enumerate}

Now, for $Z_{g}$ to be $1$, in generation $g+1$ the effective
zone of uncertainty must not reach the nonnegative area.
Therefore, $d_{g+1}$ must not be positive.
Let $H(g+1,g+2) = i$ be such a history.
Let $i'$ be constructed as follows, given $i$ and its predecessors
$h$ and $k$:

\begin{enumerate}
\item Let initial colors be the same as under $i$, except
that cell $d_{g-1}+r$ is black. (The position of $d_{g-1}$ can
be determined, given $k$.)
\item Let the life of all cells in the effective zone of uncertainty of
$i$ be determined as under $i$.
\item Let cell $d_{g-1}+r-1$ live. Thus, since the effective
zone of uncertainty of $i$ stays in the negative area, all cells
in this zone will face the same black/white decisions. Also, any
cells between $d_{g+1}$ and $d_{g-1}+r-2$ must, if they die, become white.
\item Let cell $d_{g-1}+r$ die.
\item Let cell $d_{g-1}+r+1$ live. Thus, cell $d_{g-1}+r$ will become white.
\item Let all black/white decisions in the effective
zone of uncertainty of $i$ be determined just as under $i$.
\end{enumerate}

In this generation, cells $d_{g+1}$ through $d_{g-1}+r-2$ can live or
die without affecting the inclusion of a history in $i'$.
Note that the only additional specification for
$i'$, as opposed to $i$, is the life or death of three particular cells.

Thus,

\begin{eqnarray}
\label{cb3}
P(H(g+1,g+2) \in i' \vert\\
H(g,g+1) \in h', H(g-1,g) = k') \ge\\
(\alpha_3)^3 P(H(g+1,g+2) = i \vert\\
H(g,g+1) = h,H(g-1,g) = k)
\end{eqnarray}

Note that $i$ can be recovered, given $i'$, because all
decisions in the effective zone of uncertainty of $i$ are the same.
Also note that conditions after $i'$ are the same
as after $i$. Thus, we have

\begin{eqnarray}
\label{cb4}
P(X_{g+1} = 1 \vert\\
H(g-1,g) \in k', H(g,g+1) \in h', H(g+1,g+2) \in i') =\\
P(Z_{g} = 1 \vert \\
H(g-1,g) = k, H(g,g+1) = h, H(g+1,g+2) = i)
\end{eqnarray}

Since $i'$ starts with exactly one nonnegative cell, at position
$r$ or greater, (\ref{b1}) holds.

Combining (\ref{c2}), (\ref{c3}),
and (\ref{cb3}), we have (\ref{c0}) holding
with $\kappa = {(\alpha_3)^{4r+10} \over 2^{r+1}}$.

Since there is a different $k'$, $h'$, $i'$ for each different $k$,
$h$, $i$, we have

\begin{eqnarray}
P(\exists g, X_{g+1} = 1, C_{g} = 3) \ge\\
\sum_{g,k,h,i} P(X_{g+1} = 1, C_{g} = 3 \vert\\
H(g-1,g) \in k',H(g,g+1) \in h', H(g+1,g+2) \in i')\\
P(H(g-1,g) \in k',H(g,g+1) \in h',
H(g+1,g+2 ) \in i') \ge\\
\sum_{g,k,h,i} P(Z_{g} = 1, C_{g} = 3 \vert\\
H(g-1,g) = k, H(g,g+1) = h, H(g+1,g+2) = i)\\
\kappa P(H(g-1,g) = k, H(g,g+1) = h,
H(g+1,g+2 ) = i) =\\
P(\exists g, Z_{g} = 1, C_{g} = 3)
\end{eqnarray}

Thus, by our case hypothesis, we have

\begin{equation}
P(\exists g, X_{g+1} = 1,C_{g} = 3) \ge
{\kappa \over 4} {\alpha_1 \over 2} P(\exists g, Z_{g} = 2)
\end{equation}

Case IV. (\ref {opossum}) is true with $k$ set to 4.
This case can be handled almost exactly the same as Case III.
The only difference between the two is that, in Case IV,
cell $b_{g-1}$ is not alive in generation
$g-1$. Under $k'$, this cell lives; and in this case,
if $k$ is reconstructed from $k'$, it is assumed
that this cell is not alive. Since, under $k$, the nearest
living cell to the right of $b_{g-1}$ is white, all black/white
decisions to the left of $b_{g-1}$ are the same for $k'$ as for
$k$.
\rule {2mm}{3mm}

\begin{lem}
\label{mix4}
If there is positive probability that, given any finitely
describable initial conditions, the zone of uncertainty
will expand arbitrarily far to the left (right) only, there is positive
probability $\alpha$ that, given a zone of uncertainty
consisting of two, three or four, contiguous black cells:
\begin{enumerate}
\item The effective zone of uncertainty will expand arbitrarily
far to the left, never again going to the right
of the position of the original black cells.
\item The effective zone of uncertainty will never contain less
than two black cells.
\end{enumerate}
\end{lem}

{\sl Proof.} Let $\alpha_1$ be the smallest probability that
any cell will live; and $\alpha_2$ the largest.

By Lemma \ref{l2}, if under any initial conditions there is positive
probability the zone of uncertainty will expand arbitrarily far to the
left only, there is positive probability that under these conditions
the left border of the effective zone of uncertainty will expand
arbitrarily far to the left, and the right border recede arbitrarily
far to the left.

By Corollary \ref{mix0}, if there are initial conditions under
which there is positive probability of the effective
zone of uncertainty behaving as above, then, for any $\epsilon > 0$,
there are initial conditions $I_{\epsilon}$ under which there is
probability $1 - \epsilon$ of it behaving as above.

Now, suppose that under $I_{\epsilon}$ there is probability
$1$ that the zone of uncertainty will eventually contain
one cell. Then there is probability $1 - \alpha_2$ that this
zone will eventually disappear. If $\epsilon$ is small
enough, this is a contradiction.

Therefore, at least for small enough $\epsilon$, under conditions
$I_{\epsilon}$, there is positive probability that the zone will
not only behave as above, but always contain at least two cells.
Let $\gamma$ be one such probability, for any particular conditions
$I_{\epsilon}$. Let $c$ be the cell on the right border
of $I_{\epsilon}$.

Now, under $I_{\epsilon}$, there must be some
$n$, such that that there is probablity at least ${\gamma \over 2}$
the zone of uncertainty will never reach cell $c+n$.
Let $m$ be the length of the zone of uncertainty under $I_{\epsilon}$,
plus $n$.

The proof is completed by noting that if there are two black
cells in the zone of uncertainty, there is probability at least
$\alpha_1 (1-\alpha_2)^{m} \alpha_1$ that the right black cell lives,
its $m$ neighbors on the right die, and the next white cell lives.
Given this, there is probability ${1 \over 2}^m$
that the cells that die form the pattern of $I_{\epsilon}$, with
$n$ white cells to the right. Finally, given this pattern that is
just like that of $I_{\epsilon}$, except for one black cell $c$, $n$
units to the right, there is probablity at least ${\gamma \over 2} \alpha_1
(1 - \alpha_2)^2$ that $c$ dies, its two neighbors live, and
the rest of the zone does not ever reach cell $c$.

Thus, there is probability at least $\alpha = {\gamma \over 2} (\alpha_1)^3
(1-\alpha_2)^{m+2} {1 \over 2}^m$ of
events transpiring as desired.
\rule {2mm}{3mm}

The main theorem now follows.

\begin{thm} [The Double Glider Theorem]
\label{main}
Let $G$ be a simple cellular game of radius $r$, with left/
right symmetry. Then, under $G$, with finite initial conditions,
the probability that the zone of uncertainty will extend
arbitrarily far in one direction only is zero.
\end{thm}

{\sl Proof.}
Suppose that under $G$, under any finite initial conditions,
there is positive probability of the zone of uncertainty
extending arbitrarily far to the left (or right) only. Without
loss of generality, since $G$ is symmetric, let us say the left.

Then, by Lemma \ref{l2}, there is positive probability that
both left and right effective borders of the zone of uncertainty
will move arbitarily far to the left. Since this refers
to a countable Boolean combination of finite histories, in which
no finite history determines membership,
Corollary \ref{mix0} can be applied. That is, for any $\epsilon > 0$,
there are fixed initial conditions $I_{\epsilon}$ such that, given these
initial conditions, the probability of this happening
is greater than $1 - \epsilon$.

Since this is true for any $\epsilon > 0$, let us assume
that $\epsilon < {\alpha_1 \over 2}$, where $\alpha_1$ is
the smallest probability, under $G$, that any cell will
stay alive.

Also, without loss of generality, let the rightmost black cell, under
$I_{\epsilon}$, be regarded as cell $0$. Thus, there is
probability at least $1 - \epsilon$ that, in some generation $g$,
the rightmost cell in the effective zone of uncertainty will
be at a nonnegative position, and in all subsequent generations
at a negative one.

Thus, $I_{\epsilon}$ satisfies the conditions for Lemma \ref{mix2a}.
That is, there is a constant $\gamma$ such that if, under $I_{\epsilon}$,
this crossover does occur, the probability it does so in a generation
in which, at the beginning of the generation, there is only one nonnegative
black cell (and that cell is at position $r$ or greater)
is at least $\gamma$. This $\gamma$ is not dependent on any
other characteristics of $I_{\epsilon}$, but only on $G$.

Let $X_g$ be $1$ if:
\begin{enumerate}
\item At the beginning of generation $g$, there is only one
nonnegative black cell.
\item In generations $g+1$ and later, the effective zone
of uncertainty stays out of the nonnegative area. That is,
it no longer contains nonnegative cells.
\end{enumerate}
Let $X_g$ be $0$ otherwise.

Thus, given initial conditions $I_{\epsilon}$, we can say that

\begin{eqnarray}
\label{m1a}
\gamma (1 - \epsilon) <
\sum_g P(X_g = 1,X_k = 0 \forall k < g) = \\
\sum_{g,h} P(H(1,g) = h)
P(X_g = 1,
X_k = 0 \forall k < g \vert H(1,g) = h)
\end{eqnarray}

Note that if $X_k$, with $k < g$ is  $1$, $X_g$ must be $0$; that
is, the effective zone of uncertainty can leave the nonnegative area
for the last time in only one generation. Thus, the left
side of (\ref{m1a}) becomes

\begin{equation}
\label{m1b}
\sum_{g,h} P(H(1,g) = h) P(X_g = 1 \vert H(1,g) = h)
\end{equation}

Separating out the effects of the next generation,
we get

\begin{eqnarray}
\label{m1c}
\sum_{g,h,h'} P(H(1,g) = h) P(H(g,g+1) = h' \vert H(1,g) =
h)\\
P(X_g = 1 \vert H(1,g) = h,H(g,g+1)=h')
\end{eqnarray}

Now, for it to be possible that $X_g$ be $1$, the cell history in
generations $1$ through $g-1$ must meet certain conditions. That
is, at the beginning of generation $g$ there must be only one black
nonnegative cell, at position $r$ or greater; in other words, $F_1(H(1,g))$
must be $1$. In addition, the history of generation $g$ must meet certain
requirements. That is, in generation $g+1$ the zone of
uncertainty must contain only negative cells; in other words,
$F_2(H(g,g+1))$ must be $1$. Thus, (\ref {m1c}) becomes

\begin{eqnarray}
\label{m1}
\sum_{g,h,h'} P(H(1,g) = h) F_1(H(1,g))\\
P(H(g,g+1) = h' \vert H(1,g) = h) \\
F_2(H(g,g+1))
P(X_g = 1 \vert H(1,g) = h,H(g,g+1)=h')
\end{eqnarray}

Now, given initial conditions $I_{\epsilon}$, the probability that the
zone of uncertainty does {\sl not} extend arbitrarily far to the left
only (that is, that it extends arbitrarily far to the right, or
eventually disappears) has to be less than $\epsilon$. Since
$\epsilon$ is arbitrary,
showing that this probability must be greater than some
constant dependent only on $G$ will force a contradiction.

To show this, let $r(g)$ be the position of the rightmost
cell in the effective zone of uncertainty in generation $g$. Furthermore,
let $p$ be the probability that the zone extends
arbitarily far to the right, or eventually disappears.
Then $p$ is larger than the probability that one domain in the
middle of the zone of uncertainty grows arbitrarily large in
both directions. This, in turn, is larger than the probability
that, for some generation $g$ all the conditions below hold:

\begin{enumerate}
\item
In generation $g$, there is only one nonnegative black cell $c$,
at position $r$ or greater.
\item
In generation $g+1$, there are two, three or four nonnegative
black cells, both next to each other, and both at
positions $r$ or greater.
\item
The white domain $D$ which in generation $g$ is
between cell $c$ and all other black cells, grows
arbitrarily large in both directions.
\item
In generations $g+2$ and later, either the leftmost living cell of $D$
is at position $0$ or less, or the leftmost cell in $D$ is
at position $0$ or less, and the leftmost living cell after that
is white. That is, the left effective border of $D$ is always
at position $0$ or less.
\item
In generations $g+2$ and later, either the rightmost living cell of $D$
is at position $r-1$ or less, or the rightmost cell in $D$ is
at position $r-1$ or less, and the rightmost living cell after that
is white. That is, the right effective border of $D$ is always
at position $r-1$ or less.
\item
In generation $g+2$ and after, there are always more than
two black cells to the right of $D$; that is, at positions
$r$ or greater.
\end{enumerate}

That is, a white domain $D$ develops in generation $g$, and the
two ``gliders'' on each side of $D$ in that generation
fly apart, and never touch. The right glider, after generation
$g$, always contains at least two black cells; and both
gliders continue to exist forever.

Now, we examine the probability of these events happening.
Let $Y_g$ be $1$ if the above events are satisfied for generation
$g$, and $0$ otherwise.

Thus, the probability that the zone of uncertainty grows
arbitrarily large in both directions is greater than

\begin{equation}
\label{m2}
\sum_{g,h} P(H(1,g) = h) P(Y_g = 1, Y_k = 0 \forall k < g \vert
H(1,g) = h)
\end{equation}

Now, $Y_g$ and $Y_k$, with $k < g$, cannot both be $1$. The
reason for this is that for $Y_k$ to be true, there
must be exactly one black nonnegative cell in generation
$k$, and never again. Thus, (\ref{m2}) is equivalent to

\begin{equation}
\label{m3a}
\sum_{g,h} P(H(1,g) = h) P(Y_g = 1 \vert H(1,g) = h)
\end{equation}

or, separating out the effects of generation $g$, we have

\begin{eqnarray}
\label{m3b}
\sum_{g,h,h'} P(H(1,g) = h) P(H(g,g+1) = h' \vert H(1,g) = h)\\
P(Y_g = 1 \vert H(1,g) = h,H(g,g+1)=h')
\end{eqnarray}

For $Y_g$ to be $1$, the cell history in
generations $1$ through $g-1$ must meet the same conditions that
enable $X_g$ to be $1$; that is, in generation $g$ there must be
only one black nonnegative cell, at position $r$ or greater.
In addition, the history of generation $g$ must
meet certain requirements. That is, in generation $g+1$ there
must be two or more black nonnegative cells, both next to each
other, and both in positions $r$ or greater; that is,
$F_3(H(g,g+1))$ must be $1$. Thus, (\ref {m1c}) becomes

\begin{eqnarray}
\label{m3c}
\sum_{g,h,h'} P(H(1,g) = h) F_1(H(1,g))\\
P(H(g,g+1) = h' \vert H(1,g) = h)\\
F_3(H(g,g+1))
P(Y_g = 1 \vert H(1,g) = h,H(g,g+1)=h')
\end{eqnarray}

By Lemma \ref{mix2}, for every $1$-generation history $h_2$
such that $F_2(h_2)$ $= 1$, there is a constant $\beta$
depending only on $G$, and a $1$-generation history
$h_3$ such that

\begin{enumerate}
\item
$F_3(h_3) = 1$.
\item
Initial conditions are the same as under $h_2$.
\item
For any previous
history (starting at generation $0$) $h$, we have

\begin{eqnarray}
\label{m4a}
P(H(g,g+1) = h_3 \vert H(1,g) = h) \ge\\
\beta P(H(g,g+1) = h_2 \vert H(1,g) = h)
\end{eqnarray}
\item
At the end of generation $g$, given history
$h_3$, the negative black cells are exactly
the same as those at the end of $g$ given
$h_2$.

Furthermore, for no two $h_2$ will this $h_3$ be the same.
\end {enumerate}

Thus, (\ref {m3c}) is greater than

\begin{eqnarray}
\label{m5}
\sum_{g,h,h'} P(H(1,g) = h) F_1(H(1,g))\\
\beta P(H(g,g+1) = h' \vert H(1,g) = h) \\
F_2(H(g,g+1))
P(Y_g = 1 \vert H(1,g) = h,H(g,g+1)=h')
\end{eqnarray}

For $Y_g$ to be true, the white domain $D$ must, from that
point on, include at least cells $0$ through $r$.
Therefore, by Theorem \ref{newest},
in all infinite histories for which $Y_g$ is $1$, and
all finite histories in which the possibility of $Y_g$ remaining
$1$ stays open, the actions of cells on the two sides
of $D$ remain independent of each other. Hence, these
actions can be considered separately, as if we were dealing
with two different games.
Thus, at the beginning of generation $g+2$,
the probability that behavior on both
sides of $D$ will be appropriate is the product of the probabilities
of appropriate behavior on each side.

The probability that the behavior on the left side is appropriate
is the same as the probability that behavior on the left side
would be appropriate if, at this point, the negative black
cells were exactly the same as they are now, but there were
no nonnegative black cells.

Similarly, the probability that behavior on the right side
is appropriate is just the probability that all behavior
is appropriate, if the zone of uncertainty consisted
only of two, three or four contiguous black cells. By Lemma \ref{mix4},
this probability is at least $\alpha$.

Note that this is where the left-right symmetry of $G$
comes in; that is, the probability a symmetric zone
of uncertainty will glide arbitrarily far to the left
only must be the same as the probability it will glide
arbitrarily far to the left only.

Thus, (\ref {m5}) becomes

\begin{eqnarray}
\label{m6}
\sum_{g,h,h'} P(H(1,g) = h) F_1(H(1,g))\\
\beta P(H(g,g+1) = h' \vert H(1,g) = h) \\
F_2(H(g,g+1))
P(X_g = 1 \vert H(1,g) = h,H(g,g+1)=h') \alpha
\end{eqnarray}

This sum is less than the probability, given initial conditions
$I_{\epsilon}$, that the zone of
uncertainty will expand arbitarily far in both directions;
however, by comparison to (\ref{m1a}) through (\ref{m1}), it
is seen to be greater than $\beta \alpha \gamma (1 - \epsilon)$,
with $\alpha$, $\beta$, and $\gamma$ depending only on the
game, not on the initial conditions.
If $\epsilon$ is small enough, this contradicts
the assumption that, given these conditions, this probability
must be less than $\epsilon$.
\rule {2mm}{3mm}

\section{Standard Restricted Initial Conditions}
\label{stanrec}
It may be useful to consider another form of finitely
describable initial conditions, defined as follows:

\begin{defn}
{\bf Standard restricted initial conditions} are conditions such
that there is a rightmost black cell, and a leftmost white cell.
\end{defn}

In other words, under standard restricted initial conditions,
an infinite black domain is followed, left to right, by
none, two, or any other even number of finite domains (of
alternate colors), followed by an infinite white domain.

The zone of uncertainty is defined similarly as for
finitely describable initial conditions.

\begin{defn}
\label{zone2}
Under standard restricted initial conditions,
the\- {\bf zone of uncertainty} consists those
finite domains (if any), in between the two infinite domains.
\end{defn}

In some respects, the behavior of cellular games under
these conditions is easier to analyze.
That is, if there are finitely many black cells
there is always positive probability that all black cells die out.
This essentially ends the course of the game; thus, it makes it more
awkward to discuss the long-term behavior of a system. Under standard
restricted initial conditions, however, the two infinite domains
cannot merge, and cells of each color will always be present.

Behavior under standard restricted initial conditions can
be delineated as follows:

\begin{thm}
\label{t}
Let $G$ be a simple cellular game.
Then, under standard restricted initial conditions, one,
but not both, of
the two statements below hold:

\begin{enumerate}
\item The zone of uncertainty will, almost always, become empty
infinitely many times.
\item It will, almost always, become empty only finitely many
times.
\end{enumerate}
\end{thm}

{\sl Proof.} Suppose $G$ is such that, when the zone of uncertainty
is empty, there is positive probability $p$ it is for the
last time. Then the probability that it will reach minimal
size infinitely many times is

\begin{equation}
\label{ts1}
\lim_{n \rightarrow \infty} (1-p)^n = 0
\end{equation}
\rule {2mm}{3mm}

\begin{defn} A {\bf clumping process} is
a simple cellular game in which, under standard restricted
initial conditions, the zone of uncertainty almost always becomes
empty infinitely many times.
\end{defn}

\begin{defn}
Let a simple cellular game in which this zone, almost always,
becomes empty only finitely many times be called a {\bf mixing process.}
\end{defn}

Now, there is another kind of symmetry which may be
applied to cellular games; namely, they may be black/white
symmetric, as well as left/right.

We examine clumping processes which have both symmetries.
We show that if $G$ is a clumping process with both
such symmetries, each cell will change color infinitely
many times. To do this, we use a theorem which can be applied to
all symmetric, one-dimensional random walks with the Markov property.
In this theorem, we
show that the walker will {\sl cross} any position
infinitely many times. (In the ``usual'' walk,
in which the walker can only move one unit at a
time, this means that the walker will {\sl visit} every position infinitely
many times.)

\begin{thm}
Let $M$ be a one-dimensional random walk with the Markov
property. Let $X(t)$ be the position of that walk at time $t$.
Let $P_{0,1}$ equal $p_0 > 0$, $P_{0,k}$ equal $P_{i,i+k} \forall i$, and
$P_{0,k}$ equal $P_{0,-k}$ for all $k$. Then, for any $n$,
any $g$, and any value of $X(g)$, the quantity
$P( \exists h, h > g, X(h) < n)$ equals $P(\exists h, h > g, X(h) > n)$,
and they both equal $1$. That is, this random walk will almost always
cross every position infinitely many times.
\end{thm}

{\sl Proof.}
First, the probability that the $X(g)$ will stay bounded is $0$.
That is, suppose it were not. Then, there would be some $n$
such that

\begin{equation}
\label{ts2}
P(n = \limsup_{k \rightarrow \infty} \vert X(k) \vert) > 0
\end{equation}

However, we know $P_{-n,-n-1} = P_{n,n+1} = p_0 > 0 \forall n$.
Therefore, if the walk reaches position
$n$ ($-n$) infinitely often, it will almost always reach position
$n+1$ ($-n-1$) infinitely often.

We now show that the probability that there
are infinitely many $k$, such that $X(k)$ is not the same sign
as $X(k+1)$, is $1$.

Let a sequence $\{C_i\}$ with each $C_i \in \{-1,1\}$,
and integer sequences $\{k_i\}$ and
$\{n_i\}$, be constructed as follows: By the above discussion,
we know that, with probability $1$, there must eventually
be a $k$ for which $\vert X(k) \vert \ge 2$.
Let $k_1$ be the first $k$ for which this is true, and let $n_1 =
X(k_1)$. Let $C_1$ be $1$ if $X(k_1) \ge 2$, and $-1$ if $X(k_1)
\le -2$.

Given $C_{i-1}$, $k_{i-1}$, and $n_{i-1}$, such that $X(k_i) =
n_i$, let $C_{i}$, $k_{i}$, and $n_{i}$ be constructed as
follows. Let $k_i$ be the first $k$ such that $\vert X(k_i) -
n_{i-1} \vert \ge n_{i-1}$; note that there will
almost always be such a $k_i$. Let $n_i = X(k_i)$, and let $C_i =
1$ if $n_i \ge 2 n_{i-1}$, and $-1$ if $n_i \le 2 n_{i-1}$. Thus,
if $C_i$ is a different sign from $C_{i-1}$, then $X(k_i)$ will be a
different sign from $X(k_{i-1})$.

Now, since $P_{0,-k} = P_{0,k} = P_{n,n+k} \forall n,k$
the probability that each $C_i$ is the
same sign as the previous is ${1 \over 2}$. Since
the each $C_i$ is independent of all others, they will, therefore,
almost always change sign infinitely many times.

The same argument can be used to show that, for any $c$,
$X(k) - c$ will change sign infinitely often, and hence
that any point will be crossed infinitely many times.
\rule{2mm}{3mm}

\begin{cor}
Let $G$ be a clumping process with both left/right
and black/white symmetry. Let $G$ evolve under standard
restricted initial conditions. Then, under $G$, each cell will,
almost always, change color infinitely many times.
\end{cor}

{\sl Proof.} Let $X(i)$ be the position of the leftmost
cell in the white domain, the $i$th time the zone of
uncertainty is empty. Then we know there will, almost
always, be infinitely many $X(i)$. Since cellular
game evolution is independent of exact location, $P_{0,k} =
P_{n,n+k} \forall n,k$. Since $G$ is symmetric
in both senses, $P_{0,-k}$ will equal $P_{0,k}$ for all $k$.

Now, let $\alpha$ be the smallest probability that
any cell lives, and $\beta$ the largest. By definition,
they are both positive. Let the zone of uncertainty
be empty in generation $g$ for the $i$th time. There
is probability at least $\alpha (1 - \beta) \alpha$ that
cell $X(i) - 1$ lives, cell $X(i)$ dies, and cell
$X(i) + 1$ lives. Given these events, there is probability
${1 \over 2}$ that cell $X(i)$ becomes white in
the next generation, thus ensuring that $X(i+1) = X(i) + 1$.
Thus $P_{X(i),X(i)+1}$, and hence $P_{0,1}$ and $P_{0,-1}$
must be positive. Therefore the process $X(0), X(1), \ldots,
X(n), \ldots$ satisfies the requirements of the above theorem.
\rule{2mm}{3mm}

Similar results, however, have not yet been obtained for
mixing processes. That is, we cannot show that for
mixing processes with both left/right and black/white
symmetry, evolving under standard restricted initial
conditions, the zone of uncertainty will, almost always,
expand arbitrarily far in both directions.

As shown before, there cannot,
under these conditions, be a ``gli\-der'' with two
domains of the {\sl same} color on each side of it.
This does not automatically imply that there
cannot be a ``glider'' with two domains of {\sl different}
colors on each side of it. However, the one fact
does suggest the other, which is here presented as a conjecture.

\newtheorem{conj}[defn]{Conjecture}
\begin{conj}
Let $G$ be a simple cellular game with both left/\-right and
black/white symmetry. Then, under standard restricted
initial conditions, the probability that the zone of
uncertainty will expand arbitrarily far in one direction
only is $0$.
\end{conj}

Note that if this conjecture is true, it can be shown
that under both finite and standard restricted initial
conditions, no finite domain $D$ (with probability $1$) will
grow arbitrarily large. This would be done by considering
the two areas between $D$ and the infinite domains on the left and
right to be gliders. Since $D$ will grow arbitrarily large,
each of these gliders could be shown not to be affected by what happens
on the other side of $D$. They could thus be considered
to be ``gliding'' arbitrarily far in one direction, between
two infinite domains. By Theorem \ref{main} (The
Double Glider Theorem), this is
not possible if the two domains are the same color;
and, if the above conjecture is true, this would not be possible
if the two domains are different colors.

\section{Examples}
\label{examples}
At this point, one may ask if either mixing processes or
clumping processes exist. Computer simulations suggest
that both kinds of behavior are indeed possible.

The experiments described in this chapter simulate one-dimensional
simple games of radius $1$. In these games,
the life probability of a cell is one value, $p_1$, if it is the
same color as both of its neighbors, and a different
value, $p_2$, otherwise. These games are thus both left/right
and black/white symmetric. Let such games be called
``join/mix'' processes.

Using the definition of simple cellular game, these processes can be
specified more formally as follows:
\begin{itemize}
\item There is one cell for each integer, or each integer mod $k$.
\item In each generation, each cell is either white or black.
\item If a cell is the same color as both of its neighbors,
its probability of living in that generation is $p_1 > 0$. Otherwise,
its probability of living is $p_2 > 0$.
\item If a cell lives in a generation $g$, it keeps its color
in generation $g+1$.
\item If a cell dies in generation $g$, its color in
generation $g+1$ is either that of its nearest living neighbor
to the left, or to the right, with a $50\%$ probability of each.
\item If, in generation $g$, a cell has no living neighbors
on each side, it has a $50\%$ probability of assuming either
color in generation $g+1$.
\end{itemize}

In computer experiments, games of this type are run on
a circular lattice of cells. Initially, two black
domains are placed in a mostly white area. Figure \ref {square} shows
how results vary as $p_1$ and $p_2$ vary. That is, if $p_1$
is high, there seems to be little noise at the borders
between domains. In such cases, $p_2$ determines the rate
of domain movement. If, on the other hand, $p_1$ is low
and $p_2$ high, the noise between domains seems to grow
so fast it quickly takes over the ring. If $p_1$ and
$p_2$ are both low, the asymptotic behavior of the process
is not readily apparent. However, the resemblance to
natural structures is noticeable.

\begin{defn}
The join/mix game such that $p_1 = 0.85$ and $p_2 = 0.15$
is called the the {\bf Join or Die Process}.
\end{defn}

The process is given this name because a cell must join; that
is, be the same color as both of its neighbors, or else
it is very likely to die. Computer simulations suggest that
the Join or Die process is, in fact, a clumping process. That
is, the area of ``noise'' between two large domains appears to stay,
quite small most of the time. We thus conjecture:

\begin{conj}
The Join or Die process is a clumping process.
That is, if it evolves under standard restricted initial conditions,
the zone of uncertainty will almost always become empty infinitely
many times.
\end{conj}

Now, consider what happens, under the Join or Die or other
clumping processes, to ``normal'' or ``almost all'' initial
conditions. Let us suppose that average domain size will, almost
always, grow arbitrarily large. Thus, after many generations,
most cells in any given section of the lattice would, most
likely, be in extremely large domains; and a visual depiction of
this section would show large domains, with a noisy boundary
between them (consisting of small domains, many containing no
living cells). The noisy boundary between two such large domains
would, therefore, move in some sort of symmetric random walk; and
it might be unlikely that the noise in the boundary would grow to
significant size, compared to the domains it bordered.

Thus, the evolution of such a process might be very similar to
that of a process in which the size of the ``noise'' between
domains stayed bounded. Let us suppose, without loss of
generality, that the size of the ``noise'' stayed at one cell.
Let us describe such a model (which is {\sl not} a cellular game)
as follows:

\begin{itemize}
\item
There is one cell for each integer.

\item
Each cell, at each time, is in either a black, white, or gray state.

\item
Gray domains, which may be no more than one cell wide, are called
``particles.'' Particles separate black and white domains,
which alternate.

\item
Particles move either to the right or left, in accordance with
some symmetric random walk.

\item
If two particles meet or cross, then two white domains have
absorbed a black domain (or two black domains a white one). Thus,
these two particles, which represent the boundaries between two
domains, disappear.
\end{itemize}

This is, exactly, a stochastic process discovered by Erd\H os and
Ney \cite{erdos} and called the {\sl annihilating particle
model}. And, computer simulations do, indeed, show apparent
similarities of behavior. These similarities suggest that
study of one subject may shed light on the other.

Another join/mix game is the Mixing Process.

\begin{defn}
The join/mix game such that $p_1 = 0.15$ and $p_2 = 0.85$
is called the the {\bf Mixing Process}.
\end{defn}

That is, the probabilities are exactly reversed from those
used for the Join or Die process. As this process evolves,
computer experiments suggest that the ``noise'' between
two large domains is likely to grow with time.

\begin{conj}
\label{cmix}
The Mixing Process is a mixing process.
That is, if it evolves
under standard restricted initial conditions, the zone
of uncertainty will almost always grow arbitrarily large.
\end{conj}

\end{thesis}

\appendix
\chapter {Computer Experiments}
\label {appendix1}
All computer experiments were done in Turbo Pascal, Version 4.0,
using the built-in pseudorandom number generator. Source code is
available from {\tt levine@symcom.math.uiuc.edu}.

The program simulating the modified Arthur-\-Pack\-ard-\-Ro\-gers
mo\-del, with Stag Hunt parameters, is {\tt cg2.pas}. Note that in this
program all strategies are mixed; that is, there is a small
probability of actions other than those called for by the pure
strategy.

The simulations of zero-depth, one-round models are as follows:
In Section \ref{CGM}: The Cloud Process, {\tt cloud.pas}, the
Prisoner's Dilemma, {\tt prisoner.pas}, the Stag Hunt (first
version), {\tt
stag.pas}, and the Stag Hunt (second version), {\tt stag2.pas}.
In Section \ref{TSS}: The square of different join/mix
processes, {\tt square.pas}, the Join or Die Process, {\tt
jd.pas}, the annihilating particle model, {\tt apm.pas}, and
the Mixing Process, {\tt mix.pas}.

\chapter{The Prisoner's Dilemma}
\label {appris}
The Prisoner's Dilemma is a two-person game in which two types of
moves are possible: cooperate, and defect. This game models the
options of two prisoners held in separate cells for the same
crime, who are being pressured to confess to that crime.

If both prisoners keep silent -- that is, they cooperate with each other
-- they will both get a small sentence for a lesser crime. If
they both talk -- that is, they both defect -- they both get the
standard sentence. If one talks and the other does not, the one
that kept silent gets a very severe sentence and the other goes
free. Thus, Prisoner's Dilemma is a game in which a player's
reward for defecting, while the other player cooperates, is
highest. Next highest is the reward for mutual cooperation; then,
the reward for mutual defection. Lowest of all is the reward for
cooperating while the other player defects.

The Prisoner's Dilemma can also be generalized to three-person games.
For more information on the Prisoner's Dilemma, see \cite {axelrod}
and \cite{poundstone}.

\chapter{The Arthur-Packard-Rogers Model}
\label{Rogers}

The computer experiments presented in \ref{APR} use a model very
similar to the one described in \cite{rogers}.

That is, there exists a
circular ring, or doubly infinite lattice, of cells $C$. Associated
with each cell $c$, in each round $i$ of each
generation $g$, are:

\begin{itemize}
\item A move variable $m_{c,i,g}$
from some finite alphabet $\Sigma$ of $k$
characters.

\item A strategy variable $S_{c,g}$.
This is a table, in which entries
are from $\Sigma$. If strategies are of depth $d$ and radius
$r$ (that is, moves of the $r$ nearest neighbors of a cell,
up to $d$ rounds back,
are taken into account), then this table contains $k^{d(2r+1)}$ entries.
There are hence $k^{k^{d(2r+1)}}$ possible strategies.
Note that strategies do not change in a generation, but they do
take into account rounds in previous generations.
In computer experiments, move and strategy variables are
initialized with the aid of a pseudorandom number generator.

A finite number of {\sl mixed}, that is, stochastic, strategies may also
be implemented; that is, strategies in which, given at least one game
history, there is positive probability of a cell making two different
moves. For example, a mixed strategy for Prisoner's Dilemma would be
to cooperate $95\%$ of the time, and defect the other $5\%$. If a given
game allows $k$ moves, and $k'$ mixed courses of action, there are
$(k+k')^{k^{d(2r+1)}}$ possible strategies. Again, mixed strategies, and
all other stochastic actions, are implemented with the aid of
a generator.

\item A reward, or payoff, variable $W_{c,i,g}$. This variable starts out
at $0$ in the first round of each generation, and its change in each round
measures the success of a cell in that round.
\end{itemize}

Changes to the reward variable are determined by a matrix $G$.
This matrix defines
the game and does not change during its course. That is, if a
cellular game has radius $r$, and $i > 1$,

\begin{equation}
\label{first}
W_{c,i,g} = W_{c,i-1,g} + G[m_{c-r,i,g}, \ldots, m_{c,i,g},
\ldots, m_{c+r,i,g}]
\end{equation}

An example of a game matrix is this table for a Prisoner's
Dilemma game: That is,
if ``D'' is defect, and ``C'' is cooperate: $G[CDC] = 100, $ $G[CDD] =$ $
G[DDC] = 70, $ $G[CCC] = 60, $ $G[DDD] = 40,
$ $G[DCC] = G[CCD]$ $ = 30, G[DCD] = 0$.
For this game, $k = 2$ (that is, there are two possible moves,
cooperate or defect); and $r = 1$ (only the moves of the {\sl nearest}
neighbors of a cell affect its reward variable).

In the Arthur-\-Pack\-ard-\-Ro\-gers mo\-del, a fixed number of rounds
$R$ (e.g., $150$ rounds),
is regarded as constituting a generation. After each generation, cell
strategies change, as follows:

\begin{itemize}
\item The probability of a cell ``living'' into the next generation, is
an increasing function of the size of its reward variable. Usually the
reward matrix contains only positive entries, and life probability is
proportional to the size of the reward variable of a cell.

\item A live cell keeps its strategy in the next generation.

\item A cell that does not live is given a
new strategy in the next generation. This strategy is chosen as follows:
\end{itemize}

\begin{itemize}
\item New entries in the strategy table are taken from corresponding
entries in either one of the two {\sl parent} cells (the nearest
living neighbors of a cell
on each side). The new strategy table can contain elements
from both parent cells ({\sl crossover}, Definition \ref {crossover})
or only from one parent (no crossover). The exact details of how such a
selection is carried is part of the {\sl genetic algorithm} used
in the program. For a discussion of genetic algorithms, see
\cite{goldberg}.
Note, however, that all such algorithms are symmetric between the left and
right parent; and that if a cell has no living neighbors on either side,
all strategy possibilities are equally likely.

\item After the basic new strategy is chosen, each table entry is
subject to {\sl mutation} (Definition \ref{mutation}).
That is, there is a small probability it may change.
\end{itemize}

\chapter{Figures}

\begin{figure}
\label{f1}
\centerline{\psfig{figure=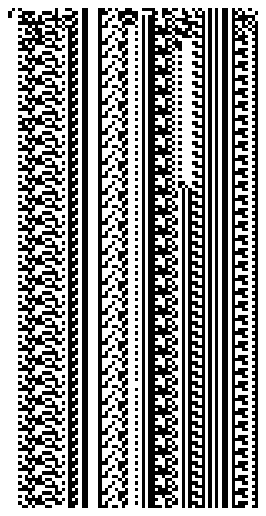,width=2.0in}}
\caption
{
Computer simulation of the Stag Hunt, a modified Arthur-Packard-
Rogers cellular game model, with $75$ cells, and $150$
generations per round. Program {\tt cg2.pas}, random seed
$824709136$, generation 1. In this program, all initial strategies
are depth $1$, but strategies of depth up to $3$ may be introduced
as the system evolves.
}\end{figure}

\begin{figure}
\centerline{\psfig{figure=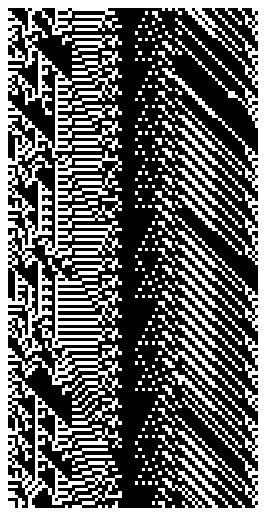,width=2.0in}}
\caption
{The same program, parameters, and seed as above, generation $27$.
Notice the rightward-moving waves of cooperative behavior, in the
right part of the display. Here some zones exhibit cellular
automaton-like triangular patterns.
}\end{figure}

\begin{figure}
\centerline{\psfig{figure=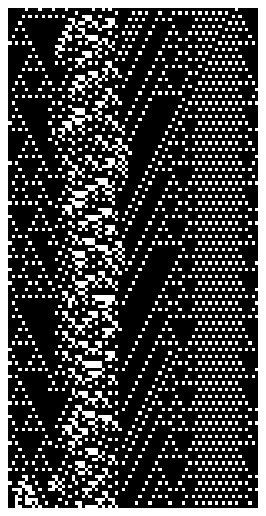,width=2.0in}}
\caption
{ Generation $139$ of this run. Cellular automaton-like
triangles predominate in this figure.
}\end{figure}

\begin{figure}
\centerline{\psfig{figure=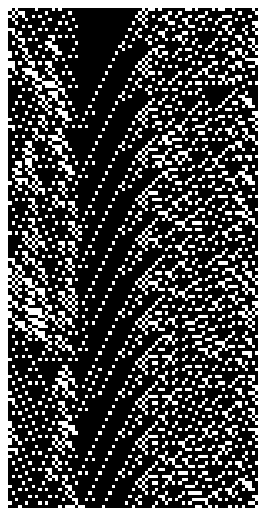,width=2.0in}}
\caption
{ Generation $165$. There are now leftward-moving waves of
cooperative behavior, in the middle of the display. }
\end{figure}

\begin{figure}
\centerline{\psfig{figure=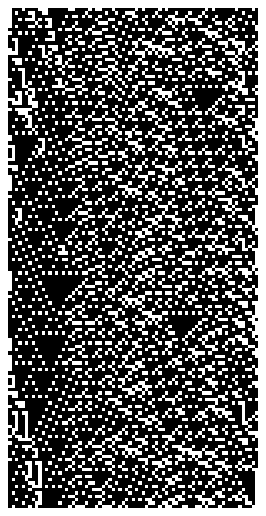,width=2.0in}}
\caption
{ Generation $305.$
}\end{figure}

\begin{figure}
\centerline{\psfig{figure=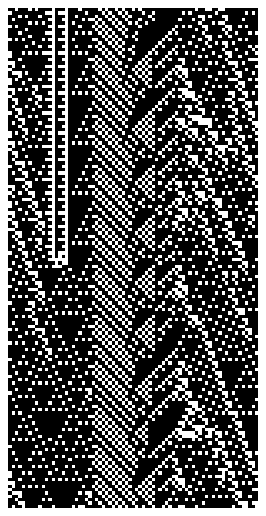,width=2.0in}}
\caption
{
Generation $483$. Cellular automaton-like triangles appear again.
}\end{figure}

\begin{figure}
\centerline{\psfig{figure=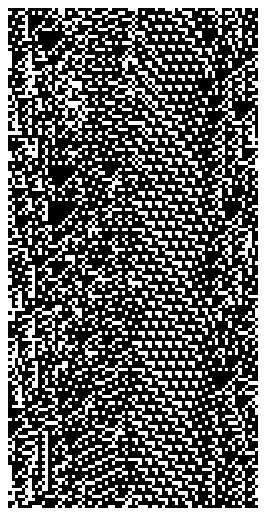,width=2.0in}}
\caption
{Generation $560$. Move behavior does not appear to have
changed much in many generations.
}\end{figure}

\begin{figure}
\centerline{\psfig{figure=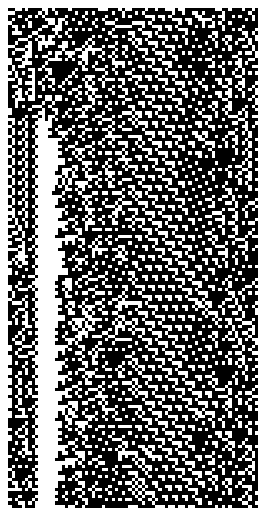,width=2.0in}}
\caption
{Generation $561$. An all-cooperate zone appears. The
next three figures show the rapid growth of this zone.
}\end{figure}

\begin{figure}
\centerline{\psfig{figure=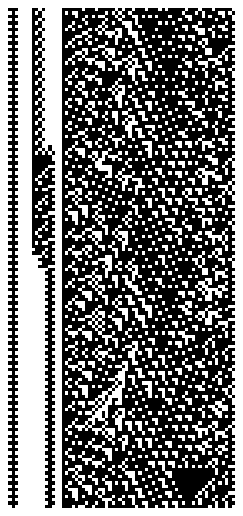,width=2.0in}}
\caption
{Generation $612$.
}\end{figure}

\begin{figure}
\centerline{\psfig{figure=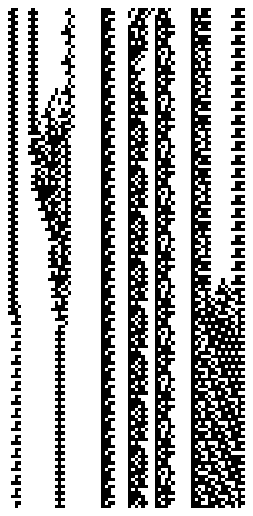,width=2.0in}}
\caption
{Generation $658$.
}\end{figure}

\begin{figure}
\centerline{\psfig{figure=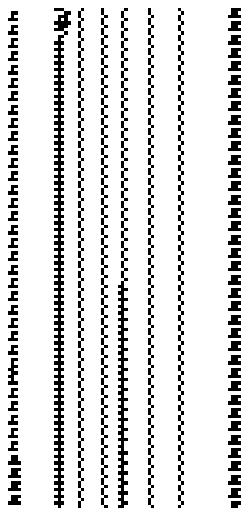,width=2.0in}}
\caption
{Generation $662$. The all-cooperate zone has almost
completely taken over the ring.
}\end{figure}

\begin{figure}
\centerline{\psfig{figure=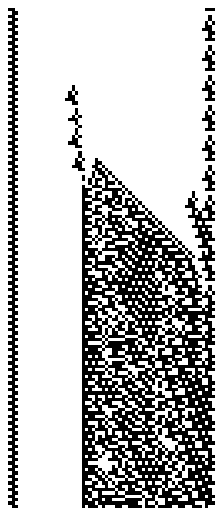,width=2.0in}}
\caption
{Generation $930$. Large all-cooperate zones have predominated
in the past several hundred generations. However, at this point,
a perturbation in strategy -- that is, an unexpected defect
move -- can set off many defect moves in other cells.
}\end{figure}

\begin{figure}
\centerline{\psfig{figure=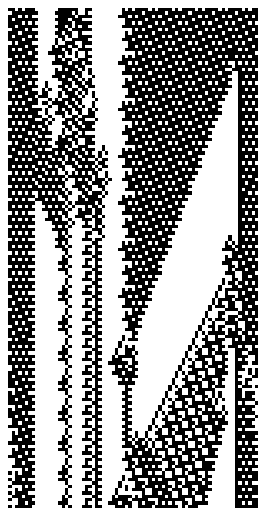,width=2.0in}}
\caption
{Generation $982$. Recovery of an all-cooperate zone.
}\end{figure}

\begin{figure}
\label{f14}
\centerline{\psfig{figure=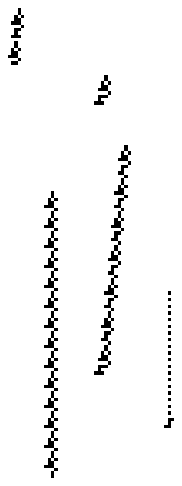,width=2.0in}}
\caption
{Generation $1262$. At this point, perturbations do not
set off much defecting behavior in other cells. That is, strategies
are no longer, ``Cooperate unless there are defectors in
the neighborhood,'' but, ``Cooperate, whatever happens.''
}\end{figure}
\begin{figure}
\clearpage

\centerline{\psfig{figure=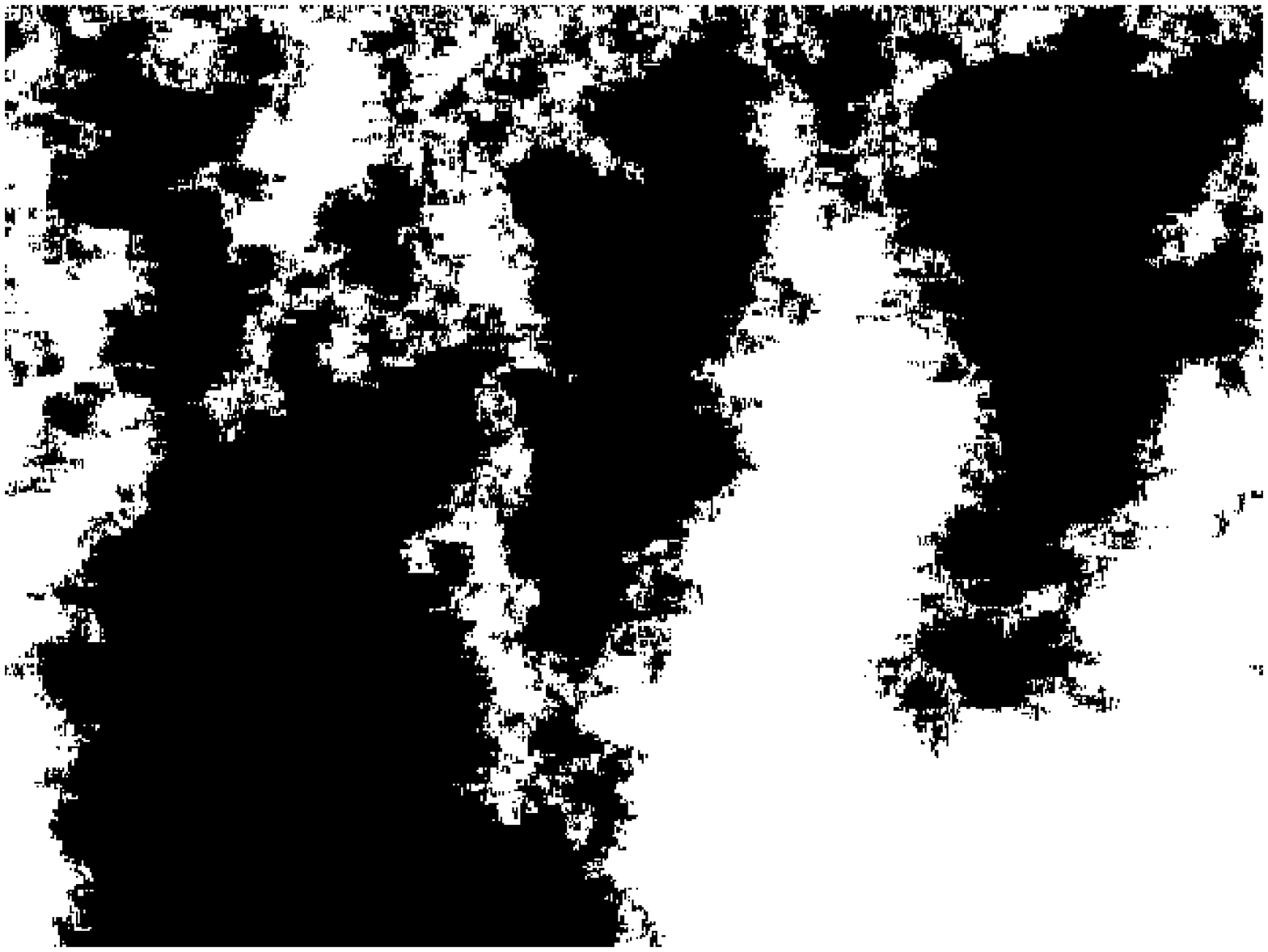,width=5.3in}}
\caption
{ \label{cl}
Computer simulation of a one-round cellular game, the Cloud
Process, on a ring of $640$ cells. The table for this game is:
$G(BBB) = G(WWW) = 0.27$, $G(BBW) = G(BWB) = G(BWW) = G(WBB)
= G(WBW) = G(WWB) = 0.53$. Program {\tt cloud.pas},
random seed $118950941$. Initial conditions were chosen with the
aid of a pseudorandom number generator.
}
\end{figure}
\clearpage

\begin{figure}
\centerline{\psfig{figure=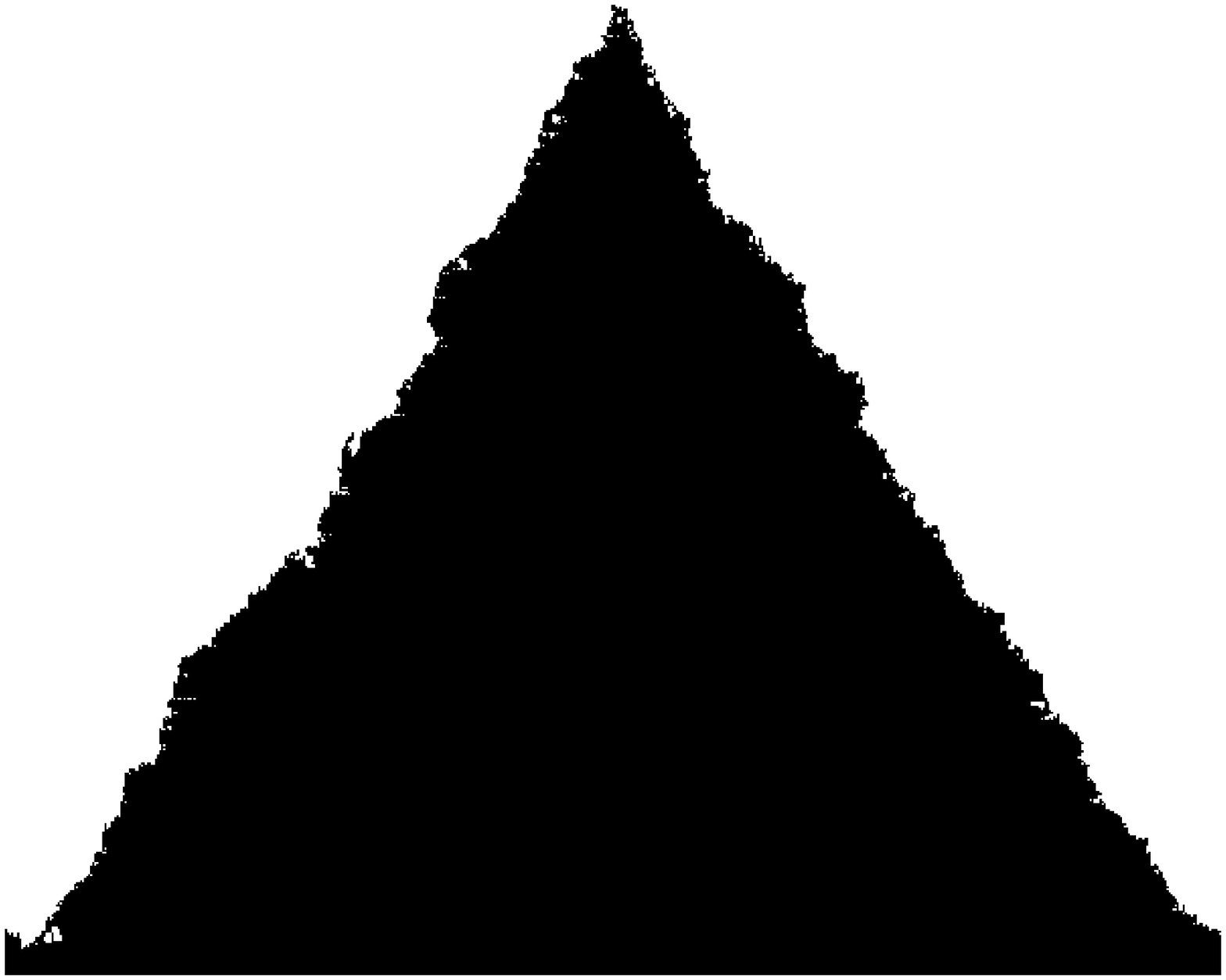,width=5.3in}}
\caption
{ \label{fpris}
Computer simulation of a one-round Prisoner's Dilemma game on a
ring of $600$ cells. Initially, two defectors are placed side-by-
side; all other cells are cooperators. (Black indicates defecting
cells, and white, cooperating.) Program {\tt prisoner.pas},
random seed $424479774$. Note that the rate of expansion of the
black domain appears roughly similar on each side, thus
suggesting an informal estimate of the expected rate.}
\end{figure}
\clearpage

\begin{figure}
\centerline{\psfig{figure=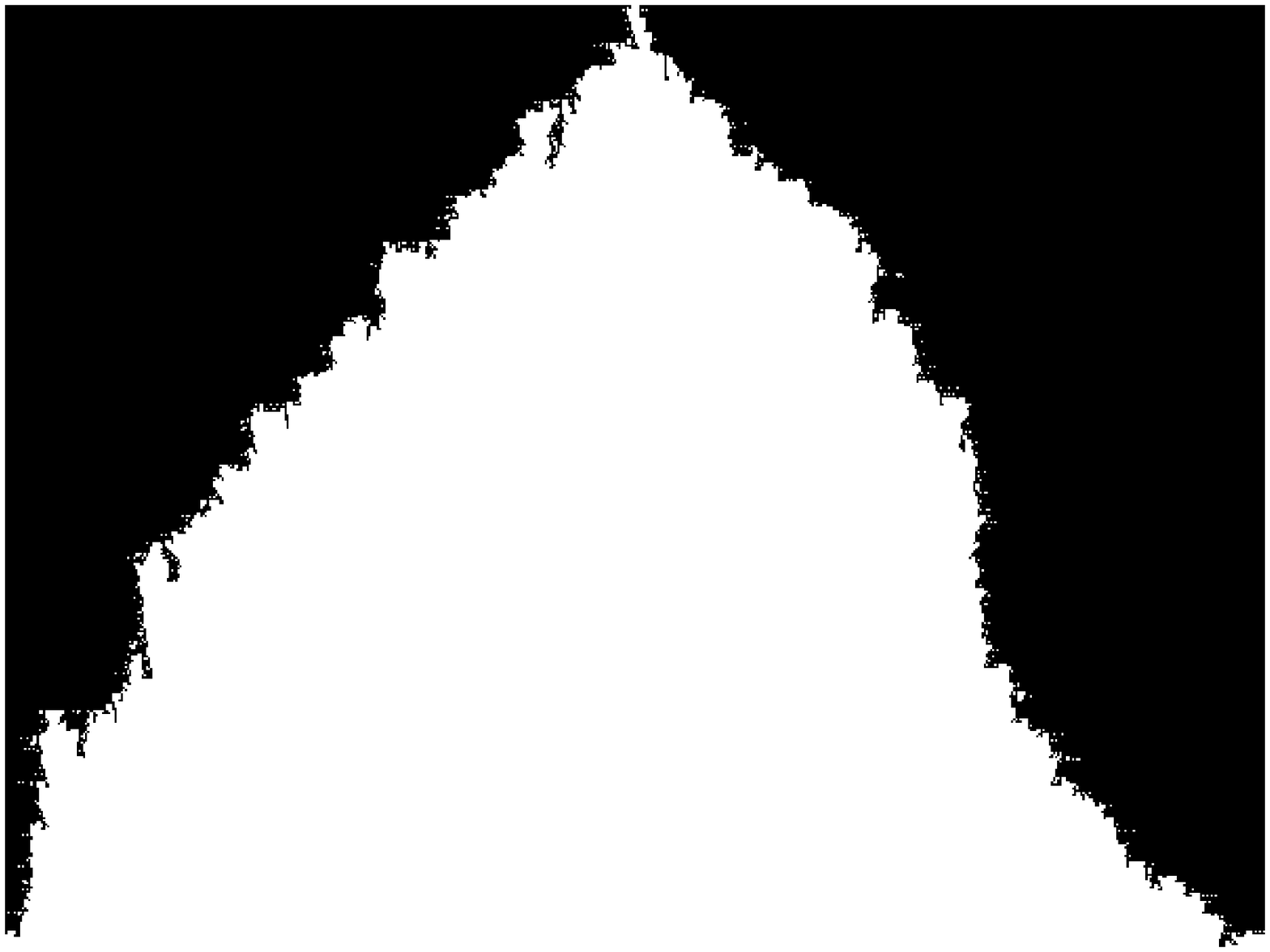,width=5.3in}}
\caption
{ \label{fstag}
Computer simulation of a zero-depth, one round per generation
Stag Hunt game on a ring of $600$ cells. Initially, four
cooperators are placed contiguously; all other cells are
defectors. Program {\tt stag.pas}, random seed $941165838$. Note
that, in this case, the rate of expansion of the white
domain appears to vary considerably. }
\end{figure}
\clearpage

\begin{figure}
\centerline{\psfig{figure=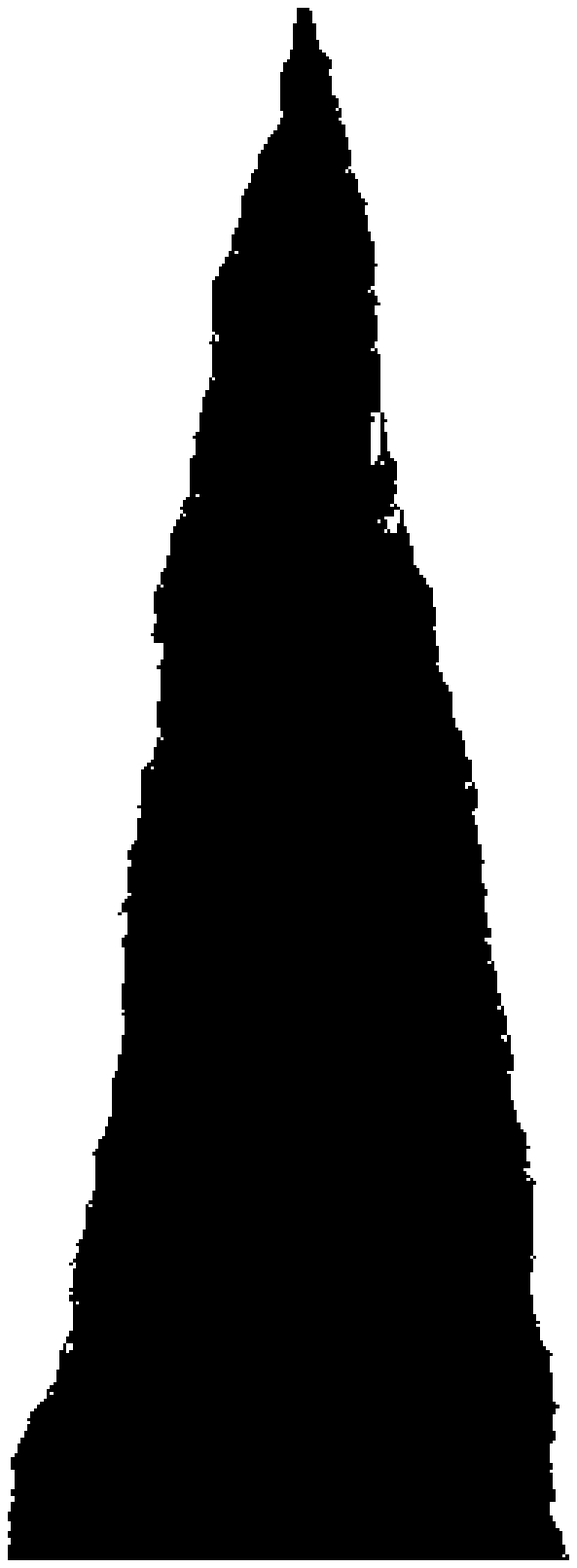,width=5.3in}}
\caption
{ \label{fstag2}
Computer simulation of a zero-depth, one round per generation
Stag Hunt game on a ring of $600$ cells. Initially, four
defectors are placed contiguously; all other cells are
cooperators. Program {\tt stag2.pas}, random seed $90049811$.}
\end{figure}
\clearpage

\begin{figure} \centerline{\psfig{figure=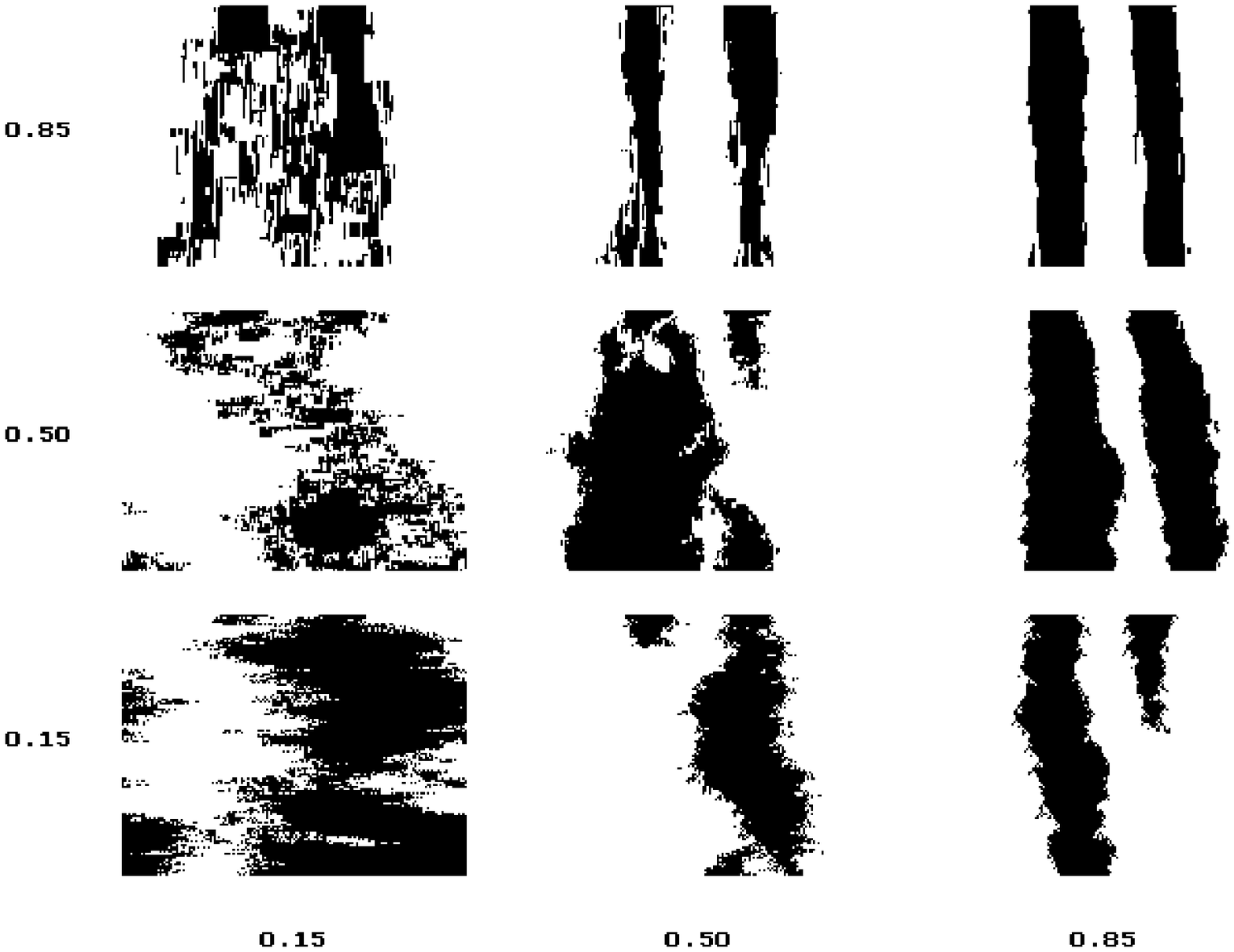,width=6in}}
\caption
{ \label{square}
Join/mix processes with various parameters. Each process is run
on a circular lattice of $165$ cells for $125$ generations. Initially,
all cells are white except for cells $48$ through $70$, and $95$
through $117$. The numbers at the bottom show the values of $p_1$;
and those on the left show the values of $p_2$. That is, both
$p_1$ and $p_2$ are set at $0.15$, $0.50$, and $0.85$. The program
used is {\tt square.pas}, seed $252644401$.}\end{figure}
\clearpage

\begin{figure}
\centerline{\psfig{figure=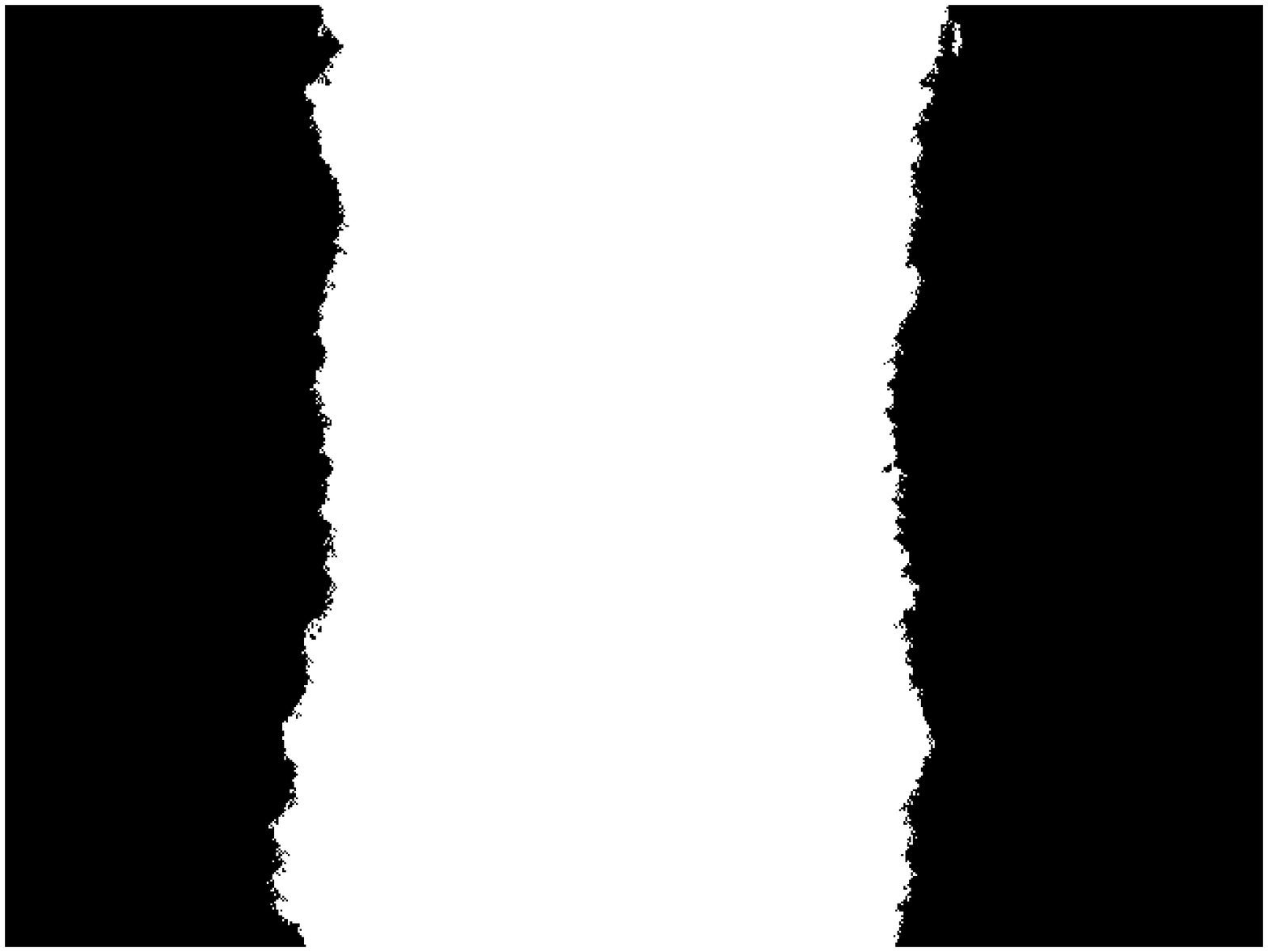,width=5.3in}}
\caption
{Computer simulation of the Join or Die
Process on a ring of $640$ cells. Initial conditions are black
for cells $0$ through $127$, and $512$ through $639$, and white
for cells $128$ through $511$. Program {\tt jd.pas}, random seed
$274535429$. }\end{figure}
\clearpage

\begin{figure} \centerline{\psfig{figure=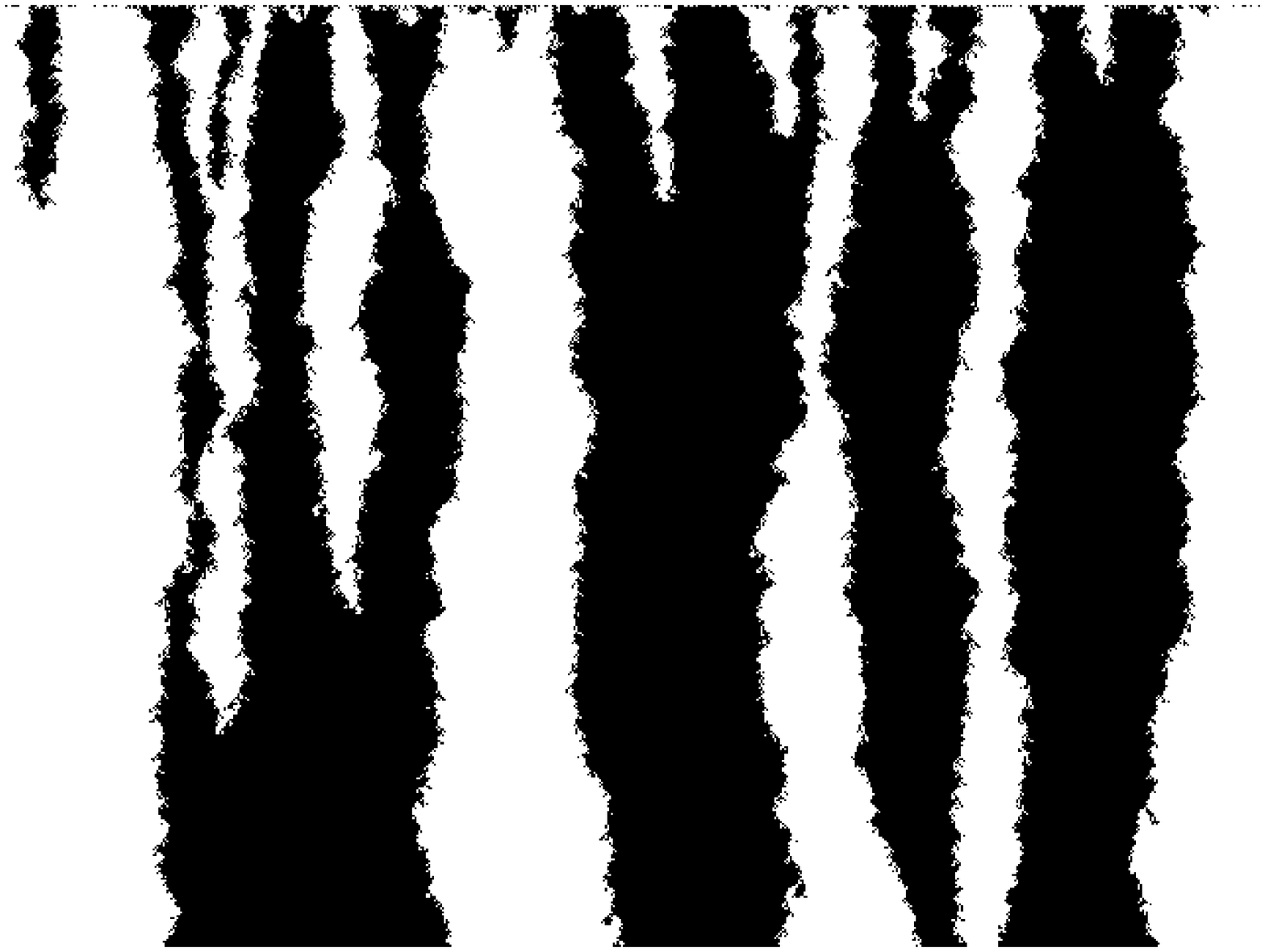,width=5.3in}}
\caption { Computer simulation of the Join or Die Process.
Initial conditions were chosen with the aid of a pseudorandom
number generator, so each cell is equally likely to be black or
white. Random seed $705238026$ is used; and the same program, and
ring size, as in the preceding figure. } \end{figure}
\clearpage

\begin{figure}
\centerline{\psfig{figure=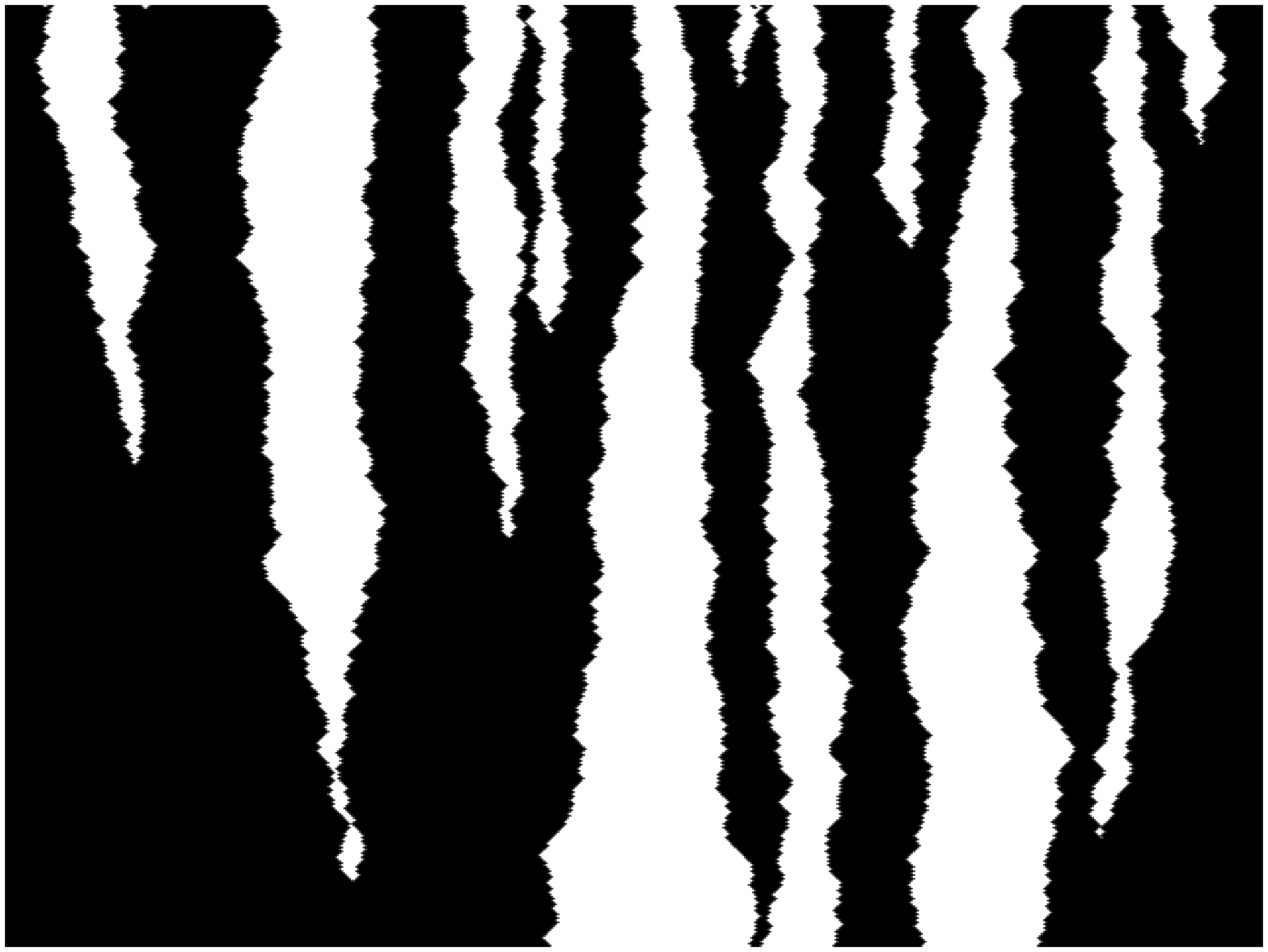,width=5.3in}}
\caption
{
The annihilating particle model, on a ring of $640$ cells, which
initially contains $28$ particles. Program {\tt apm.pas}, seed
$269093635$. Each particle executes a symmetric random walk,
having a $50\%$ probability of going either left or right in
each generation.
} \end{figure}
\clearpage

\begin{figure}
\centerline{\psfig{figure=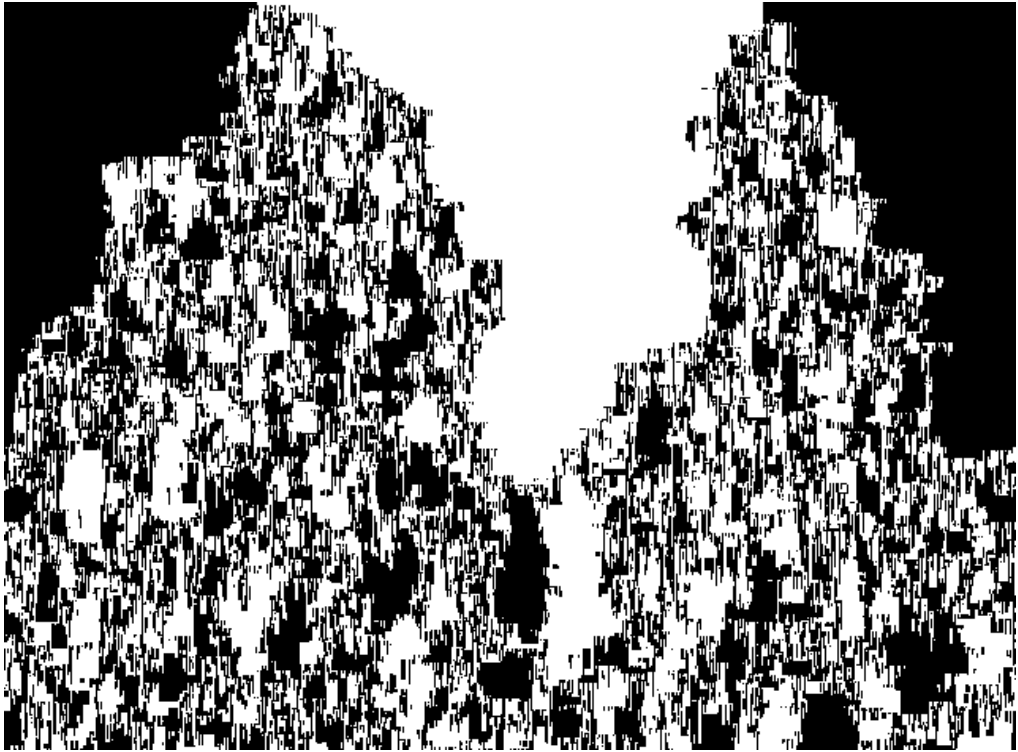,width=5.3in}}
\caption
{Computer simulation of the Mixing Process on a ring
of $640$ cells. Initial conditions are black
for cells $0$ through $127$, and $512$ through $639$, and white
for cells $128$ through $511$. Program {\tt mix.pas}, random seed
$912200719$.}\end{figure}
\clearpage

\bibliographystyle{plain}

\begin{thebibliography}{999}

\bibitem{axelrod}
Axelrod, R., {\sl The Evolution of Cooperation}, (Basic Books,
Inc., 1984).

\bibitem{berger}
Berger, R., ``The Undecidability of the Domino Problem,'' {\sl Mem. Amer.
Math. Soc.}, {\bf 66} (1966).

\bibitem{cowan}
Cowan, R., and Miller, J. H., The Santa Fe Institute,
``Life on a Lattice: The Nature of
Equilibria in Spatially Overlapping Games,'' unpublished.

\bibitem{cowan2} Cowan, R. and Miller, J. H., ``Economic Life on a
Lattice: Some Game Theoretic Results,'' Santa Fe Institute Working
Paper 90-10, 1990.

\bibitem{erdos}
Erd\H os, P., and Ney, P., ``Some Problems on Random Intervals and
Annihilating Particles,'' {\sl The Annals of Probability}, {\bf 2:5}
(1974) 828-839.

\bibitem{goldberg}
Goldberg, D. E., {\sl Genetic Algorithms in Search, Optimization,
and Machine Learning}, (Addison-Wesley, 1989).

\bibitem{langton}
Langton, C. G., {\sl Artificial Life}, (Addison-Wesley, 1989).

\bibitem{levy}
Levy, Steven, {\sl Artificial Life}, (Pantheon, 1992).

\bibitem{luce}
Luce, R. D., and Raffia, H., {\sl Games and Decisions}, (Wiley,
1957; Dover, 1985).

\bibitem{mat1} Matsuo, K., and Adachi, N., ``Metastable Antagonistic
Equilibrium and Stable Cooperative Equilibrium in Distributed
Prisoner's Dilemma Game,'' Proceedings of the International
Symposium on Systems Research, Informatics and Cybernetics, Baden-Baden,
1989.

\bibitem{mat2} Matsuo, K., and Adachi, N., ``How to Attain to
Cooperative Society in Game World: The Choice of Selection Rules,''
preprint, The International Institute for Advanced Study of Social
Information Science, Fujitsu Laboratories, Ltd., Japan.

\bibitem{mat3} Matsuo, K., and Adachi, N., ``Ecological Dynamics
under Different Selection Rules in Distributed and Iterated
Prisoner's Dilemma Game,''
preprint, The International Institute for Advanced Study of Social
Information Science, Fujitsu Laboratories, Ltd., Japan.

\bibitem{miller}
Miller, J. H., ``The Evolution of Automata in the Repeated Prisoner's
Dilemma,'' Essay in {\sl Ph.D. Dissertation, the University of Michigan},
1988.

\bibitem{mitchell}
Mitchell, M., Hraber, P. T., and Crutchfield, J. P., ``Revisiting the
Edge of Chaos: Evolving Cellular Automata to Perform Computations,''
Santa Fe Institute Working Paper 93-03-014, 1993.

\bibitem{nowak}
Nowak, M. A., and May, R. M., ``Evolutionary Games and Spatial Chaos,''
{\sl Nature}, {\bf 359} (1992) 826.

\bibitem{packard}
Packard, N., The Prediction Company, personal communication (1993).

\bibitem{packard2}
Packard, N., ``Adaptation Toward the Edge of Chaos.'' In {\sl Dynamic
Patterns in Complex Systems}, pages 293-301, (World Scientific,
Singapore, 1988).

\bibitem{poundstone}
Poundstone, W., {\sl The Prisoner's Dilemma}, (Doubleday, 1993).

\bibitem{robinson}
Robinson, R. M., ``Undecidability and Nonperiodicity for Tilings of
the Plane,'' {\sl Inventiones Mathematicae}, {\bf 12} (1971) ,177-209.

\bibitem{rogers}
Rogers, K. C., ``Cellular Automata Simulations Exhibiting an
Evolutionary Increase in Complexity,'' {\sl Master's Thesis, the
University of Illinois, Department of Electrical Engineering}, (1990).

\bibitem{rudin}
Rudin, W., {\sl Real and Complex Analysis}, 3rd ed, (McGraw-Hill,
1987).

\bibitem{waldrop} Waldrop, M. M., {\sl Complexity}, (Simon and
Schuster, 1992).

\bibitem{waterman} Waterman, M. S., ``Some Applications of Information
Theory to Cellular Automata,'' {\sl Physica 10D}, (1984), 45-51.

\bibitem{wolfram}
Wolfram, S., ``Statistical Mechanics of Cellular Automata,''
{\sl Review of Modern Physics}, {\bf 55} (1983), 601-644.

\bibitem{wolfram2}
Wolfram, S., ``Universality and Complexity in Cellular
Automata,'' {\sl Physica 10D}, (1984),  1-35.

\end{thebibliography}

\begin{vita}
Lenore E. Levine was born in Brooklyn, New York. She graduated from
Regents College with a B.S., concentration in Computer Science,
in May 1988, and from the University of Illinois with a M.S. in
Mathematics in January 1991. From 1984 to 1988, she worked as
a software engineer for DPI Services, Inc., in San Jose, California.
She entered the Graduate College of the University of Illinois at
Urbana-Champaign in August 1988. While at Illinois, she held a fellowship
and a teaching assistantship.
\end{vita}

\end{document}